\documentclass[journal=ancac3,manuscript=article]{achemso}
\usepackage{amsfonts}
\usepackage{amsmath}
\usepackage{amssymb}
\usepackage{graphicx}
\usepackage{bm}
\usepackage{setspace}
\usepackage{longtable}\makeatletter
\usepackage{setspace}

\usepackage{color}

\title{Overlimiting Current and Shock Electrodialysis in Porous Media}
\author{Daosheng Deng}
\author{E. Victoria Dydek}
\altaffiliation{ Present address: Department of Surgery, Beth Israel Deaconess Medical Center, Harvard Medical School, Boston, MA 02215 and Wyss Institute of Biologically Inspired Engineering, Harvard University, Boston, MA 02215 }
\author{\textcolor{black}{Ji-Hyung Han}}
\author{Sven Schlumpberger}
\author{Ali Mani}
\altaffiliation{Present address: Department of Mechanical Engineering, Stanford University, Stanford, CA 94304 USA}
\author{\textcolor{black}{Boris Zaltzman}}
\altaffiliation{Blaustein Institutes for Desert Research, Ben-Gurion University of the Negev, Sede Boqer Campus, 84990, Israel}
\author{Martin Z. Bazant }
\affiliation{Department of Chemical Engineering, Massachusetts Institute of Technology, Cambridge, MA 02139 USA}
\alsoaffiliation{Department of Mathematics, Massachusetts Institute of Technology, Cambridge, MA 02139 USA}
\email{bazant@mit.edu}
\date{\today}

\begin{document}

\begin{abstract}
Most electrochemical processes, such as electrodialysis, are limited by diffusion, but in porous media, surface conduction and electro-osmotic flow also contribute to ionic fluxes. In this paper, we report experimental evidence for surface-driven over-limiting current (faster than diffusion) and deionization shocks (propagating salt removal) in a porous medium.
The apparatus consists of a silica glass frit (1 mm thick with 500 nm mean pore size) in an aqueous electrolyte (CuSO$_4$ or AgNO$_3$) passing ionic current from a reservoir to a cation-selective membrane (Nafion).  The current-voltage relation \textcolor{black}{of the whole system} is consistent with \textcolor{black}{a proposed theory based on} the electro-osmotic flow mechanism over a broad range of reservoir salt concentrations (0.1 mM - 1.0 M), after accounting for (Cu) electrode polarization and pH-regulated silica charge.  Above the limiting current, deionized water ($\approx 10 \mu$ $M$) can be continuously extracted from the frit, which implies the existence of a stable shock propagating against the flow, \textcolor{black}{bordering a depleted region that extends more than 0.5mm across the outlet}. The results suggest the feasibility of ``shock electrodialysis" as a new approach to water desalination and other electrochemical separations.
\end{abstract}


\section{ \textcolor{black}{ Introduction } } 

Electrochemistry is playing an increasing role in sustainable world development. Besides energy conversion and storage, electrochemical systems also provide unique capabilities for desalination and other separations.
The availability of fresh water may soon exceed that of energy as a global concern, which will require advances in water purification technologies \cite{shannon2008,humplik2011,greenlee2009,potts1981}. Water treatment is also a key challenge for energy-related industrial processes, such as hydraulic fracturing  (``fracking") for shale gas extraction~\cite{schultz_book}.
The most difficult step is the removal of dissolved salts, especially multivalent ions
~\cite{volesky1995,WanNgah2008,Yavuz2006}. Reverse osmosis (RO) \textcolor{black}{driven by a mechanical pressure} is \textcolor{black}{perhaps the most} widely used for large-scale seawater desalination, \textcolor{black}{but consumes a high amount energy, resulting in the need of an energy efficient desalination method}. Electrochemical methods, such as electrodialysis~\cite{probstein1994,sonin1968,nikonenko2010} (ED) and capacitive deionization~\cite{johnson1971,biesheuvel2009MCDI,biesheuvel2010,zhao2012,suss2012} (CD), can be attractive for brackish or wastewater treatment and for compact, portable systems.

\begin{figure}
\includegraphics[width=5in,keepaspectratio=true]{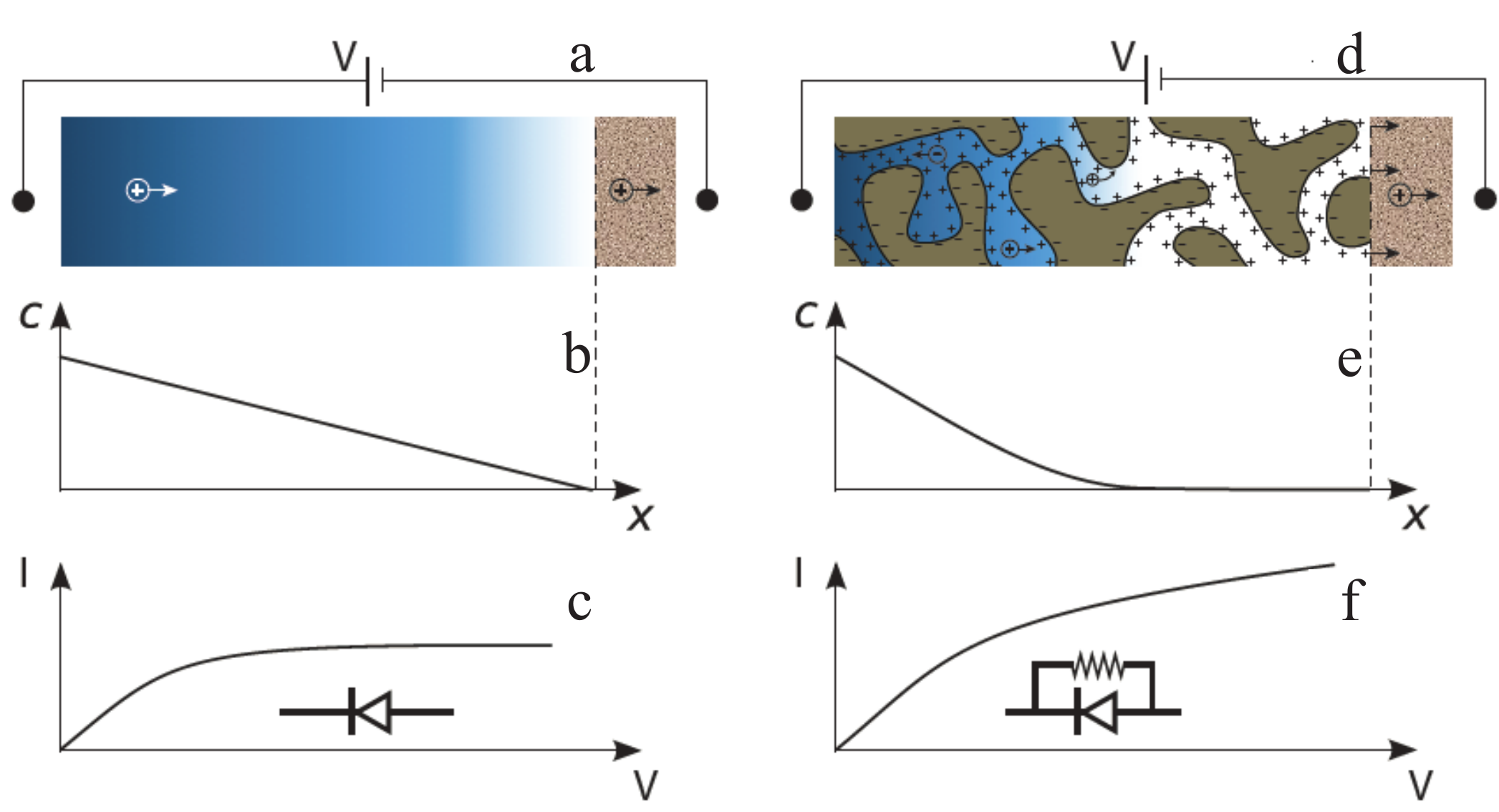}
\caption{ \textcolor{black}{ Ion concentration polarization in a bulk electrolyte (left panels) versus a porous medium (right panels). Steady current is applied from a reservoir (on the left) through an ideal membrane or other cation-selective surface (on the right). Bulk electrolyte: (a) Sketch of the bulk system; (b) salt concentration profile at the diffusion-limited current, which vanishes at the membrane; (c) current-voltage relation exhibiting limiting current, analogous to an ideal semiconductor diode; 
Porous medium: (d) Sketch of charged pores with thin double layers that control over-limiting conductance; (e) concentration profile extrapolating to zero away from the membrane under over-limiting current; (f) equivalent circuit consisting of a the bulk ideal diode in parallel with a shunt resistance for surface transport.  }
}
\label{fig:1}
\end{figure}

The rate-limiting step electrochemical separations, including ED~\cite{sonin1968,nikonenko2010} and CD~\cite{bazant2004,biesheuvel2010,suss2012}, is typically diffusion. The classical diffusion-limited current arises when (say) cations are completely removed at a membrane or electrode surface, as anions are rejected to maintain neutrality (\ref{fig:1}a).  For a dilute $z:z$ electrolyte, ambipolar diffusion leads to a linear concentration profile in steady state (\ref{fig:1}b), and the current-voltage relationship \cite{probstein1994,dydek2011},
\begin{equation}
I=I_{lim}\left[1-\exp\left(-\frac{zeV}{k_BT}\right)\right],
\label{eq:basic}
\end{equation}
 is equivalent to that of an ideal diode (\ref{fig:1}c), where
\begin{equation}
I_{lim}=\frac{2zeDc_0}{L}\textcolor{black}{A}  \label{eq:I_lim}
\end{equation}
is the diffusion-limited current, \textcolor{black}{$A$ the area of current collector}, $I$ the measured current,
 $V$  the voltage across the electrolyte, $k_B$  Boltzmann's constant, $T$  the temperature, $e$ the electron charge, $D$ the cation diffusion coefficient, $c_0$ the \textcolor{black}{reservoir} ion concentration, and $L$ the diffusion length, from the reservoir to the selective surface. Above the thermal voltage, $V \gg k_BT/e$ ($=26$mV at room temperature), the current saturates, $I \to I_{lim}$, like a diode under reverse bias.

In practice, overlimiting current (OLC), which exceeds $I_{lim}$, is often observed, and its possible origins have long been debated \cite{nikonenko2010}. For {\it bulk} transport, the consensus is that OLC can arise from chemical effects, which create more ions \cite{nikonenko2010,frilette1957,block1966,simons1979,rubinstein1984} or reduce membrane selectivity~\cite{andersen2012}, or from convection by electro-osmotic instability near the membrane ~\cite{rubinstein1988,zaltzman2007,rubinstein2008,yossifon2008}. More intriguingly, it has recently been predicted that {\it surface} transport can also sustain  OLC in a microchannel by electro-osmotic flow~\cite{yaroshchuk2011} (EOF) or surface conduction~\cite{dydek2011} (SC) along the sidewalls, depending on the aspect ratio and surface charge. The new theory~\cite{dydek2011} may explain different ion concentration polarization (ICP) phenomena observed at micro/nano-channel junctions~\cite{kim2007,yossifon2010}.

A surprising feature of microfluidic experiments \textcolor{black}{in the regime of overlimiting current} is the tendency for the depleted region to form a very sharp boundary with the bulk electrolyte~\cite{wang2005,kim2007,kim2010}, which can be understood as a shock wave in the salt concentration, propagating against the flow~\cite{mani2009,zangle2009,zangle2010}.
It has recently been predicted that stable ``deionization shocks" can also propagate in {\it porous media} at constant current~\cite{mani2011,yaroshchuk2012acis,dydek2013}, and the theory predicts steady OLC in a finite system at constant voltage (\ref{fig:1}d-f)~\cite{dydek2011,yaroshchuk2012acis,dydek2013}.

In this article, we investigate OLC experimentally in materials with sub-micrometer mean pore size. The results are consistent with theoretical predictions and reveal some basic principles of nonlinear electrokinetics in porous media.  Classical electrokinetic phenomena, such as streaming potential and electro-osmotic flow, are defined by the linear response of flow or current to a small applied voltage or pressure~\cite{probstein1994,lyklema_book_vol2}, but relatively little is known about the nonlinear response of a porous medium to a large voltage ($V \gg k_BT/e = 26$ mV at room temperature).  In contrast to recent work on induced-charge electrokinetics in polarizable media~\cite{cocis2010,large_acis},  we focus on surfaces of (nearly) fixed charge and report the first experimental evidence that surface transport can sustain OLC and deionization shocks over macroscopic distances in a porous medium.

{\color{black}

\section{  Theory }

\subsection{ Over-limiting conductance }

The classical theory of concentration polarization assumes a homogeneous bulk electrolyte~\cite{probstein1994,lyklema_book_vol2}, but there is a growing realization that new  nonlinear electrokinetic phenomena arise when the electrolyte is weakly confined by charged surfaces aligned with the applied current~\cite{mani2009,zangle2009,zangle2010,mani2011,yaroshchuk2011,yaroshchuk2011acis,yaroshchuk2012nano,yaroshchuk2012acis,dydek2011,dydek2013,rubinstein2013}.  Under strong confinement with overlapping double layers, a nanochannel or pore acts as a counterion-selective membrane, since the pore is effectively ``all surface"~\cite{heyden2005}.  Under weak confinement with thin double layers, it is well known that surface conduction plays only a small role in linear electrokinetic phenomena because the total excess surface conductivity is much smaller than the total bulk conductivity (small Dukhin number)~\cite{bikerman1940,werner2001,delgado2007}. Concentration polarization alters this picture, and surprisingly, surface-driven transport can dominate at high voltage, even with initially thin double layers.  As the bulk becomes locally depleted, the notion of excess surface conductivity (used to define the Dukhin number) becomes less useful~\cite{dydek2013}, and the key dimensionless parameter becomes the ratio of surface charge to co-ion charge per volume~\cite{mani2011,dydek2011,yaroshchuk2012nano}. 

A simple theory of surface-driven OLC in a microchannel was recently proposed by our group ~\cite{dydek2011}. The theory predicts  two new mechanisms for OLC whose relative importance is controlled by the surface charge and channel aspect ratio.  In thin or highly charged channels, the dominant mechanism is SC, while in thick or weakly charged channels, it is EOF, as long as the viscosity is also low enough for sufficiently fast flow.  In water with typical surface charges, the predicted transition from SC to EOF occurs at the scale of several microns for dilute electrolytes (mM) or tens of nanometers for concentrated electrolytes (M). For both mechanisms, the current-voltage relation is approximately linear just above the limiting current, \begin{equation}
I=I_{lim}\left[1-\exp\left(-\frac{zeV}{k_BT}\right)\right]+\sigma_{OLC}V,
\label{eq:addsurface}
\end{equation}
as if the surfaces provide a constant shunt resistance to bypass the diode-like response of ICP  (Fig. \ref{fig:1}(f)). The scalings of $\sigma_{OLC}$ with salt concentration, surface charge, and geometry are quite different for the two mechanisms, which allows them to be distinguished experimentally (below). The theory is consistent with microfluidic ICP experiments visualizing concentration and flow profiles~\cite{kim2007}, although a systematic test of the predicted scalings (below) remains to be done.

The goal of this work is to investigate to what extent these new phenomena also exist in porous media, which offer the natural   means to exploit surface-driven OLC for practical macroscopic systems. Since there is no general theory of nonlinear electrokinetic phenomena in porous media, we adapt the microchannel scaling analysis for porous media and then compare to experimental data. 

\subsection{ Mechanism 1:  Surface Conduction }

Even during strong concentration polarization, the homogenized effect of SC in porous media without flow can be rigorously described by the Leaky Membrane Model~\cite{dydek2013,yaroshchuk2012acis,mani2011,schmuck_preprint},  where the pore surface charge per volume enters the electroneutrality condition as a  fixed charge density, as in classical models of ion-exchange membranes~\cite{teorell1935,meyer1936,spiegler1971}. For one-dimensional transport over a distance $L$ from a reservoir to an ideal cation-selective surface in a dilute binary electrolyte, the exact solution of the model yields  (\ref{eq:addsurface}) with the over-limiting conductance,
\begin{equation}
\sigma_{OLC}^{SC} \approx \frac{2zeAD_m q_s}{k_BT L h_p}    \label{eq:SC}
\end{equation}
where $D_m$ is the macroscopic diffusivity, $-q_s$ is the (negative) pore surface charge per area, and $h_p=a_p^{-1}$ is the mean pore size, equal to the inverse of $a_p$, the internal area density (area/volume))~\cite{dydek2011,dydek2013}.  In the absence of flow, we have $D_m=\epsilon_p D_0/\tau_p$, where $D_0$ is the molecular diffusivity in free solution, $\epsilon_p$ the porosity, and $\tau_p$ the tortuosity.   Electro-osmotic convection in the double layers contributes a small $\sim 10\%$ correction to the surface conductivity in a dead-end channel~\cite{dydek2011}, which can be neglected in thin pores. The primary effect of electro-osmotic flow is hydrodynamic dispersion, which can lead to much larger effective $D$, as well as a second mechanism for over-limiting current.

\begin{figure}
\includegraphics[width=3.5in,keepaspectratio=true]{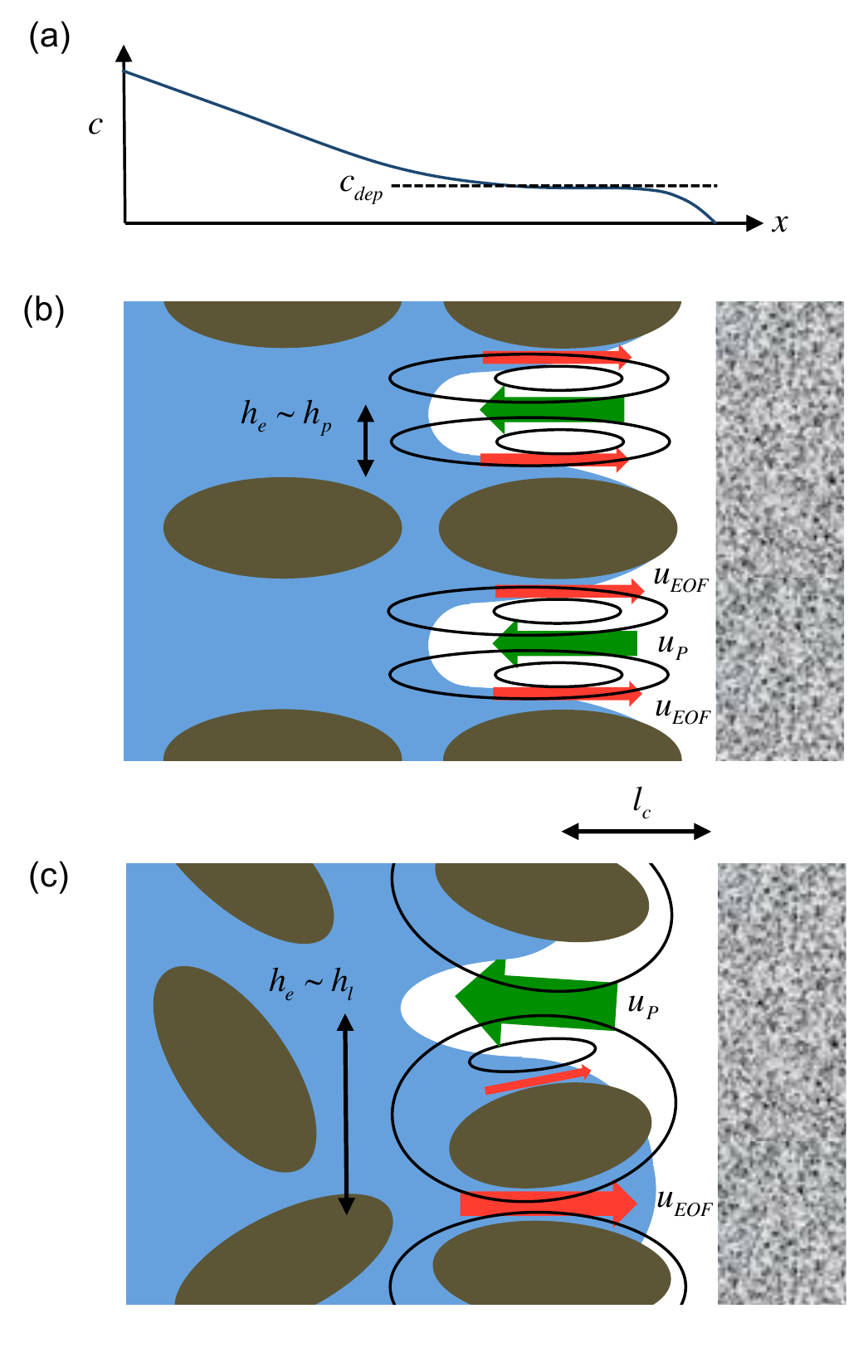}
\caption{ \textcolor{black}{ Electro-osmotic flow (EOF) mechanism for over-limiting current in a porous medium (brown matrix) pressed against an impermeable, ideally counter-ion selective membrane (on the right). Since the net flow is zero, the EOF (red arrows) driven by the large electric field in the depleted region is balanced by pressure-driven back flow (green arrows). The resulting electro-convective eddies (black loops) push salty fingers (black) toward the membrane faster than diffusion.  The eddies have transverse  size, $h_e$, and centers at a mean distance $l_c$ from the membrane.   (a) Mean salt concentration profile, indicating in the mean value $c_d$ in the depleted region.  (b)   In a regular array of equal-sized pores, the eddies are confined to parallel single-pore channels of width $h_p$, as in a straight microchannel~\cite{dydek2011,rubinstein2013}. (c) In an irregular pore network, variations in hydraulic resistance produce eddies around connected loops of characteristic width $h_l$.
}}
\label{fig:EOF}
\end{figure}

\subsection{  Mechanism 2:  Electro-osmotic Flow }

As the pore width is increased,  convection  by EOF becomes important in the bulk region and eventually dominates SC. The possibility of OLC sustained by EOF was first proposed by Yaroshchuk et al. based on a classical Taylor-Aris dispersion model that assumes EOF vortices in the depleted region are slow enough for strong transverse mixing by molecular diffusion, but it turns out that this condition is typically violated~\cite{yaroshchuk2011}.  Dydek et al. then showed that EOF can support OLC by fast electro-osmotic convection in the depleted region (Fig. \ref{fig:EOF}(a)), leading to nonuniform salt profiles with diffusion localized in a thin boundary layer near the walls~\cite{dydek2011} (\ref{fig:EOF}(b)).  Rubinstein and Zaltzman recently analyzed this new mode of dispersion in the simpler case of a neutral solute transported by a pair of vortices from a reservoir to an absorbing surface through a single microchannel by constant slip velocity on the side walls,   and they described the emergence of ``wall fingers" transitioning to spiral structures with increasing convection (Peclet number)~\cite{rubinstein2013}.

As a simple first approximation for the over-limiting conductance due to EOF in a porous medium, we adapt the microchannel scaling analysis of Dydek et al.~\cite{dydek2011}.  Electro-osmotic flow scales as $u \sim \varepsilon \zeta E /\eta$,
where the zeta potential, $\zeta \sim q_s \lambda_D/\varepsilon$, is related to the surface charge density $q_s$ using the thin diffuse-layer capacitance ($C = \varepsilon/\lambda_D$). The mean tangential electric field, $E$,  is related to the local mean current density, $j$, via $E \sim j / \sigma_b$, where $\lambda_D(c)$ is the Debye length and $\sigma_b(c)$ is the bulk conductivity, each depending on the local bulk salt concentration $c$.  Combining these equations, we obtain the EOF velocity scaling, $u \sim q_s \lambda_D j / \sigma_b \eta$.

The porous medium is pressed against an impermeable, ideally selective membrane for counter-ions (opposite to the pore surface charge). In order to ensure zero mean flow, pressure builds up near the membrane and drives a non-uniform back flow that balances the electro-osmotic forward flow and leads to hydrodynamic dispersion ~\cite{yaroshchuk2011}. In the case of a single microchannel, the sum of these flows is a vortex pair (for flat plates) or ring (for a cylindrical channel) that produces wall fingers carrying salt to the membrane, as well as internal fingers carrying depleted solution away from the membrane, faster than diffusion~\cite{dydek2011,rubinstein2013}.  For a regular array of pores (\ref{fig:EOF}(b)), similar behavior arises, and any variations in pore size or connectivity lead to  fluctuations in the wall finger flux between parallel channels. The mean transverse thickness of the eddies is thus $h_e \sim h_p$ for a regular microstructure.  

For an irregular microstructure with different pore sizes (\ref{fig:EOF}(c)), variations in hydraulic resistance lead to non-uniform pressure-driven back flow that can exceed the electro-osmotic flow in the larger pores. In that case, the mean eddy size is set by the typical width of a connected loop between nearby pores of high and low hydraulic resistance, $h_e \sim h_l$.  As the current increased and the depleted region extends across larger distances, the flow structure becomes more complicated and possibly chaotic, as in the case of electro-osmotic instability in free solution~\cite{druzgalski2013}, with ``eddy fingers" extending around many loops in the porous network.  Clearly, over-limiting conductance by EOF dispersion in porous media is quite complicated and remains to be analyzed in detail, but it turns out that the following simple derivation manages to capture the observed scalings in our experiments below. 

Consider fast electro-osmotic convection in the depleted region leading to eddy fingers of transverse thickness, $h_e$ (set by the mean size of either pores or loops), and axial length $l_c$, set by the mean distance from the membrane to the eddy centers, as shown in \ref{fig:EOF}. As in boundary-layer analysis of forced convection in a pipe~\cite{deen_book}, the convection-diffusion equation, $\vec{u}\cdot\nabla c = D_m \nabla^2 c$, then yields the scaling,
$u/l_c \sim D_m/h_e^2$. As the eddy size increases at larger over-limiting currents, the effective diffusivity $D$ incorporates porosity and tortuosity factors, as well as corrections due to microscopic hydrodynamic dispersion (such as Taylor dispersion) at a length scales smaller than the eddy size. Combining the convection-diffusion scaling with the electro-osmotic flow scalings above, we find $l_c \sim u h_e^2/D_m \sim  q_s \lambda_D j h_e^2  / \sigma_b \eta  D_m$.

Assuming the same current-voltage relation as for a single microchannel (Eq. 8 of Dydek et al. \cite{dydek2011}), the over-limiting conductance, $\sigma_{OLC} \sim 2(ze)^2 c_d D_m A  / k_B T L$, is set by the mean salt concentration $c_d$ in the depleted region (\ref{fig:EOF}(a)).  Although this region contains fingers of non-uniform salt concentration, we use the mean value $c_d$ to  define the local bulk conductivity, $\sigma_b = \varepsilon D/\lambda_D^2$ and   $\lambda_D^2 = \varepsilon k_B T /2(ze)^2 c$, respectively, as a first approximation.  All that remains then is to determine $c_d$. Numerical simulations in a straight microchannel exhibit the scaling, $c_d/c_0 \sim l_c/L$, close to the limiting current ~\cite{dydek2011}. If the same relation also holds in a porous medium for $j \sim j_{\lim} \sim 2 ze c_0 D_m/L$, then we can eliminate $l_c$ and arrive at a scaling law for the over-limiting conductance,
\begin{equation}
\sigma_{OLC}^{EOF} \sim \frac{  (2 c_0 h_e)^{4/5} q_s^{2/5} (ze)^{6/5} \varepsilon^{1/5}  D_m^{3/5} A  }{(\eta k_B T)^{2/5} L^{9/5} }
\label{eq:EOF}
\end{equation}
where there is an unknown numerical prefactor, independent of all the parameters. The best way to determine the prefactor, once the scalings are validated, is by experiment (below).

\section{  Experiment  }

\subsection{ Apparatus} 
}

The apparatus is designed to test the theoretical current-voltage relation (\ref{eq:addsurface}) for a charged porous medium and extract the over-limiting conductance.   By choosing a copper electrolytic cell~\cite{mattsson1959,brown1965,rosso2007} with known Faradaic reaction resistance $R_{F}(c_0)$ and a porous silica glass frit with known surface charge $q_s(\mbox{pH})$ in water~\cite{behrens2001,heyden2005,gentil2006}, the current-voltage relation of the frit can be isolated. {\color{black}
The dependence of $\sigma_{OLC}$ on $c_0$, which also involves $q_s$ from measurements of pH($c_0$), can be compared with theoretical formulae for the EOF and SC mechanisms, \ref{eq:SC} and \ref{eq:EOF}, respectively, having clearly distinct scalings:
\begin{equation}
\sigma_{OLC} \sim \left\{ \begin{array}{ll} 
q_s & \ \  \mbox{SC}  \\
q_s^{2/5} c_0^{4/5} &\ \ \mbox{EOF}
\end{array} \right.   \label{eq:scalings}
\end{equation}
The dominant mechanism should have a larger predicted conductance. 
}

The predicted salt depletion is directly tested by extracting deionized water from the glass frit by pressure-driven flow near the membrane interface. The flow rate is precisely controlled by a syringe pump (Harvard apparatus pump 33). The pH of the solution is measured by a pH meter (Thermo scientific, Orion PH meter), and conductivity is obtained by electrochemical impedance spectroscopy (Gamry instrument Reference 3000). 

\begin{figure}
\includegraphics[width=4in,keepaspectratio=true]{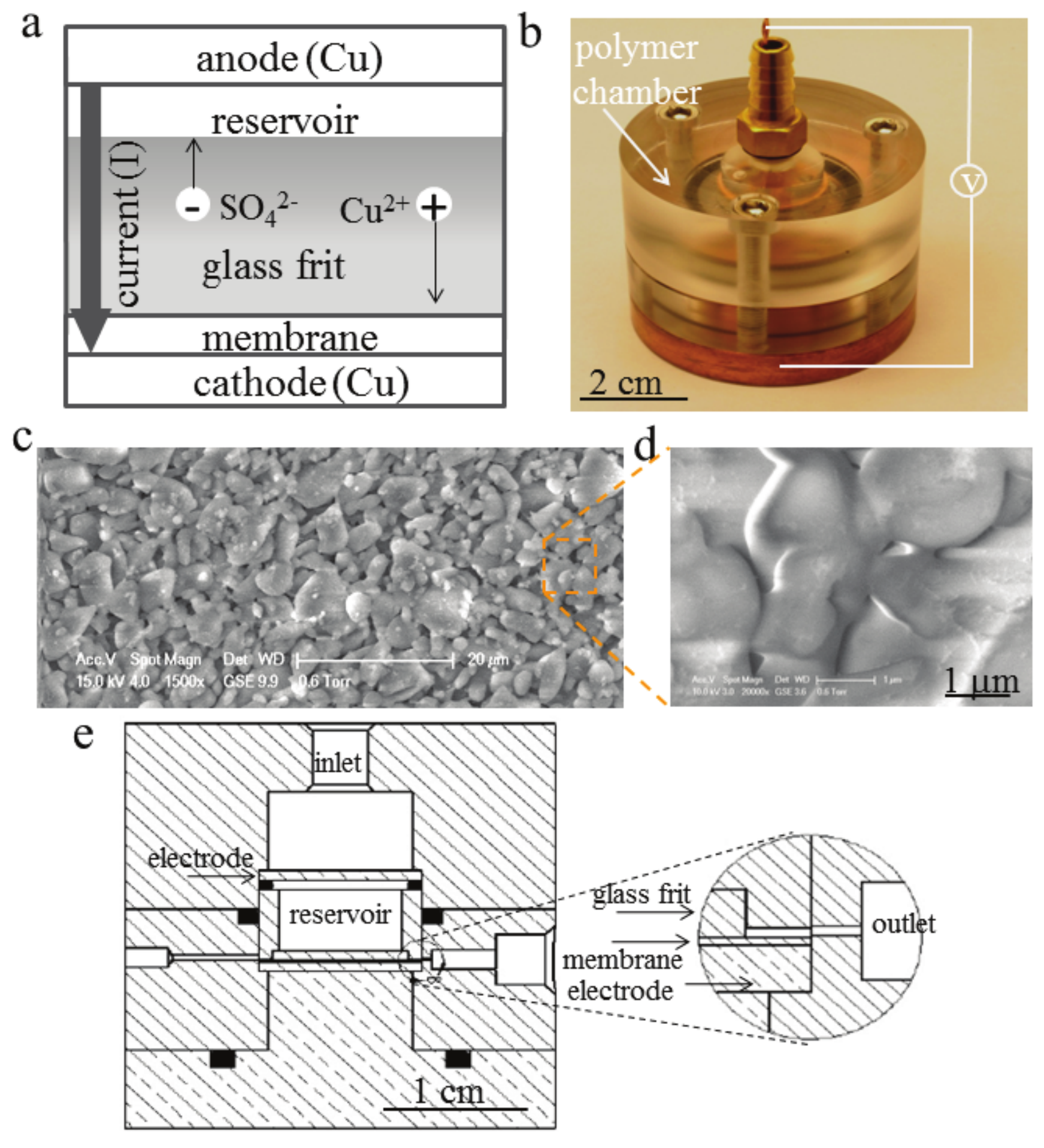}
\caption{Prototype ``button cell" for shock electrodialysis.  (a) Sketch of the frit/membrane/electrode sandwich structure (not to scale), (b) Photograph of prototype, (c) Scanning electron microscopy (SEM) image of the glass frit, showing the distribution of pores, and (d) enlarged micrographs consistent with the mean pore size around 500 nm. (e) Cross-section drawing to scale. Right: enlargement showing the radial outlet for fresh water extraction. }
\label{fig:2}
\end{figure}

The key component of the apparatus is a cylindrical silica glass frit ($L=1$ mm thick, $R=5$ mm radius) pressed with screws against a Nafion membrane, in direct contact with a copper disk cathode (\ref{fig:2}). The frit is separated from a copper disk anode by an electrolyte reservoir ($L_0=3$ mm thick).  The electrolyte is aqueous copper sulfate (CuSO$_4$), so the well-studied copper deposition/dissolution reaction~\cite{mattsson1959,brown1965,rosso2007,semichaevsky2010},
\begin{eqnarray}
\mbox{Cu}^{2+} + 2e^- &\longleftrightarrow& \mbox{Cu},  \label{eq:copper}
\end{eqnarray}
is favored at both electrodes.  Concentrated 1 M aqueous CuSO$_4$ solution is prepared by dissolving copper sulfate in the distilled water, and more dilute solutions obtained by adding distilled water. The solution is infused into the chamber, and the glass frit and membrane are immersed in the solution for several hours prior to measurements. The frit (from Adams \& Chittenden Scientific Glass) has a random microstructure of sub-micron pores, mostly $500-700$ nm wide (\ref{fig:2}c, d) with BET internal area, $a_m=1.75$ m$^2$/g, and mass density, $\rho_m=1.02$ g/cm$^3$. The pore area density, $a_p = a_m\rho_m = h_p^{-1}$, implies a mean pore size of $h_p = 557$ nm.  \textcolor{black}{As discussed below, the bare surface charge of the pores is regulated by pH, and some experiments are also done with two different chemical modifications to control the surface charge. }

The Nafion membrane serves as the cation-selective surface in \ref{fig:1}, triggering salt depletion in the glass frit. In principle, the copper cathode could play this role by itself, but inserting the membrane better mimics theoretical models~\cite{dydek2011} and possible applications to water treatment below.  
The membrane also reduces the activation overpotential at the cathode (by maintaining high cation concentration) and suppresses dendritic growth into the glass frit (due to
pressurized contact with a space-filling material). Indeed, no dendrites are observed at the cathode, and the current is stable over the range of applied voltages. \textcolor{black}{ Nafion may also help to suppress hydrogen gas evolution at the copper cathode relative to electrodeposition since Cu$^{2+}$ competes with H$^+$ in carrying the current, while the water reduction product OH$^-$ is blocked. } 

\subsection{ Electrochemical measurements } 
The current-voltage curves are measured using linear-sweep voltammetry with an electrochemical analyzer (Uniscan instruments PG581).  An optimal scan rate of 2 mV/s is chosen to attain quasi-steady response, while avoiding concentration polarization in the reservoir, which develops over several hours  at constant voltage. {\color{black} The typical time to reach OLC at roughly 0.5 V is around 5 min., which is somewhat smaller than the diffusion time across the frit, $L^2/D\approx 20$ min. Along with convection in the reservoir (see below), this minimizes reservoir salt depletion, but it leaves some transient diffusion effects in the current versus time signals, namely an initial bump or overshoot at the limiting current followed by weak oscillations around the approximately linear mean profile. Transient effects are also investigated by chronoamperometry in the Supporting Information. }

\begin{figure}
\includegraphics[width=6in]{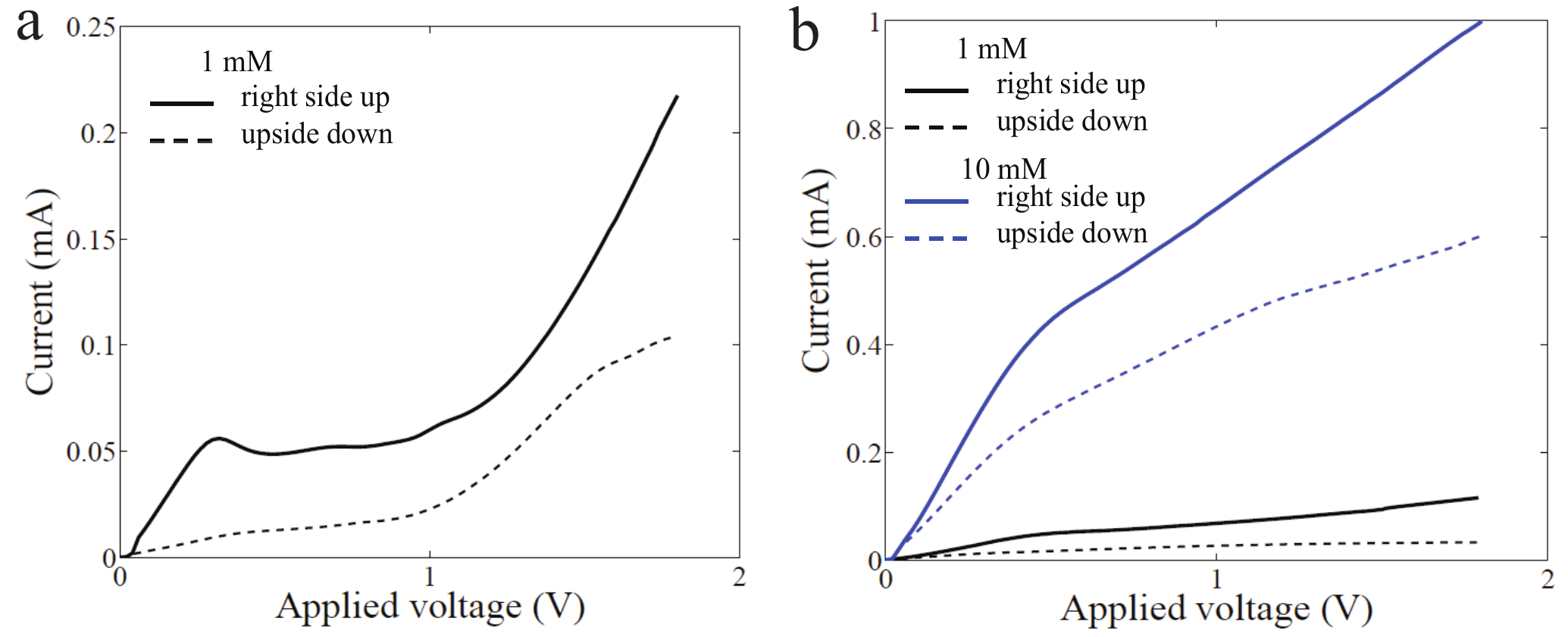}
\caption{ \textcolor{black}{ Effects of reservoir convection. (a) When the glass frit is removed, the current voltage relation shows the well-known nonlinear increase in over-limiting current due to electro-osmotic instability in free solution~\cite{rubinstein2008}. The same behavior occurs when the cell is turned upside down (cathode at the top) although the current is reduced by the absence of natural convection in the stable density gradient of copper sulfate depletion. (b) When the glass frit is inserted, the expected kinked linear current-voltage relation \ref{eq:addsurface} is observed when the cell is right side  up  (cathode at the bottom) due to vigorous mixing of the reservoir by natural convection, but the kink is suppressed and the resistance increased when the cell is upside down.   }}
\label{fig:convection}
\end{figure}

{\color{black}
A number of preliminary experiments are performed to validate the interpretation of the data below.
We first confirm (\ref{fig:convection}a) that \ref{eq:addsurface} holds with the glass frit in place, but changes to reflect the well-known nonlinear increase of OLC above a critical voltage due to electro-osmotic instability~\cite{rubinstein2008}, when the frit is removed.  Next we consider natural convection due to buoyancy forces associated with copper sulfate depletion, which have been previously observed in copper electrodeposition experiments~\cite{huth1995,rosso2007}. Copper sulfate is heavier than water, so the depleted fluid produced at the cathode tends to rise, while the enriched fluid at the anode sinks.  By repeating a number of experiments upside down, the effect of gravity becomes clear (\ref{fig:convection}).

In our apparatus, bouyancy plays a crucial role by mixing the reservoir without significantly affecting transport within the glass frit, where buoyancy forces are weak compared to viscous and electrostatic forces.  Natural convection is controlled by the Rayleigh number, $\mbox{Ra} =  (\Delta \rho/\rho) g L^3 / \nu D$, where $\Delta \rho/\rho$ is the relative fluid density variation, $g$ the gravitational acceleration, $\nu$ the kinematic viscosity, $D$ the salt diffusivity, and $L$  the characteristic length scale.  At the reservoir scale, $L = 1$mm, natural convection is strong since $\mbox{Ra} = 10^7   \Delta \rho/\rho$. When the system is right side up with the cathode producing lighter fluid at the bottom, natural convection vigorously mixes the reservoir solution via the Rayleigh-Taylor instability since the Rayleigh number is much larger than the critical value $\mbox{Ra}\approx 10^{3}$. Bouyancy thus helps to enforce a constant salt concentration boundary condition at the edge of the frit during the passage of current, as assumed in the theoretical models used to interpret the data below.  When the system is turned upside down, lighter fluid is produced at the top of the cell, leading to a stable density gradient without natural convection. This promotes slow transient diffusion into the reservoir, requiring many hours ($L^2/D$) to reach steady state and blurring the kink signifying the transition to over-limiting current in the glass frit (\ref{fig:convection}b). The experimental setup thus takes advantage of gravity to isolate the quasi-steady nonlinear current-voltage relation of the glass frit from spurious effects of transient diffusion in the reservoir.
}

\section {Results }

\subsection{ Current-voltage relation }

A typical voltammogram is shown in \ref{fig:3}a (black curve) for $c_0=1$ M. As in all prior experiments with ion-exchange membranes~\cite{nikonenko2010} and microfluidic devices~\cite{rubinstein2008,kim2010,yossifon2010},  the classical diffusion-limited behavior of \ref{eq:basic} (gray line in \ref{fig:3}), which has no free parameters and saturates at the thermal voltage $k_BT/e= 26$ mV, does not match the data.  In our experiment, we resolve this discrepancy by accounting for the electrode and reservoir polarization. A series resistance, $R_s$, is fitted to the low-voltage portion of each voltammogram by replacing the applied voltage, $V_{app}$, with
\begin{equation}
V=V_{app} - I R_s   \label{eq:Vcorr}
\end{equation}
in {\ref{eq:basic}, as in the black curve in \ref{fig:3}a.
The experiments are thus able to infer the electrokinetic response of the glass frit, which can be separated from all the other internal resistances by examining the scalings of $I_{lim}$, $R_s$, and $\sigma_{OLC}$ with salt concentration and surface charge.

\begin{figure}
\includegraphics[width=6.5in]{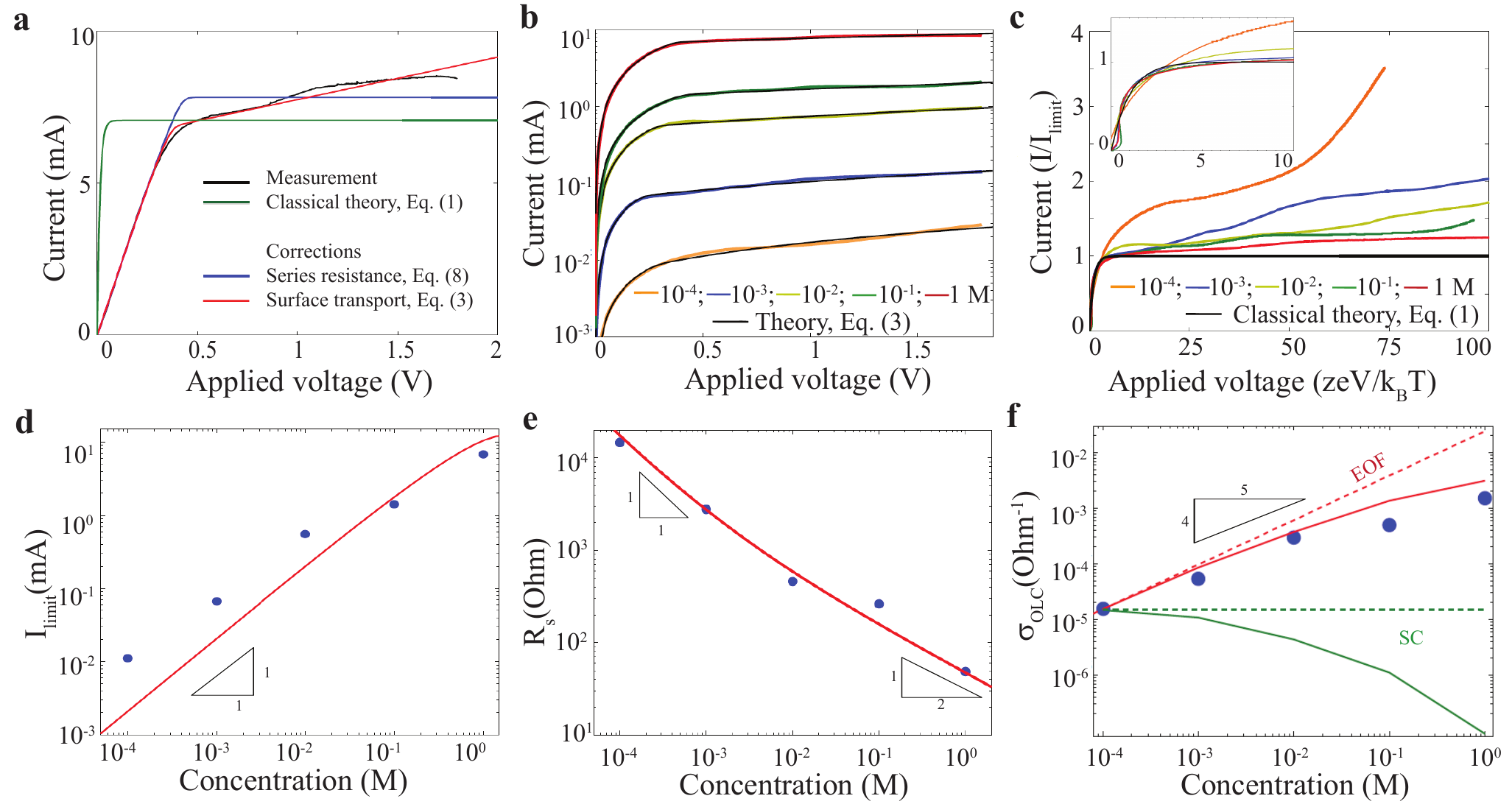}
\caption{Observation of overlimiting current in aqueous CuSO$_4$. (a) For 1.0 M concentration, the current-voltage curve is compared with the classical diffusion-limited model (\ref{eq:basic}), the extension for OLC by surface transport (\ref{eq:addsurface}), and its correction for electrode resistance (\ref{eq:Vcorr}). (b) Current-voltage data for varying initial ion concentrations with fitted curves based on the new model. (c) Dimensionless voltage-current curves with an inset showing the data collapse at lower voltages. Bottom row presents the scaling of the fitting parameters with salt concentration $c_0$ (black points): (d) limiting current $I_{lim}$, (e) series resistance $R_s$ and (f) over-limiting conductance $\sigma_{OLC}$, compared with theoretical curves (solid colors) and scalings (slopes) discussed in the text.}
\label{fig:3}
\end{figure}

After correcting for electrode polarization, the theoretical prediction of ~ \ref{eq:addsurface} provides a good fit of the data (red line in \ref{fig:3}a). By least-squares fitting of only three parameters ($I_{lim}$, $R_s$, and $\sigma_{OLC}$) for each quasi-steady voltammogram, the coefficient of determination is 99\% over a wide range of salt concentrations (\ref{fig:3}b).
The limiting current from the fitting for 1 M concentration is about 6.8 mA, and the membrane working area is the same as the cross-section area of glass frit, resulting in the calculated limiting current density approximately 8.6 mA/cm$^2$.
As shown in \ref{fig:3}c, the dimensionless current, $\tilde{I} =I/I_{lim}$, versus  voltage, $\tilde{V}=zeV/k_BT$ ($z=2$), collapses onto a single master curve,  \ref{eq:basic}, at low voltage (inset), while displaying a nearly constant over-limiting conductance, consistent with the theory~\cite{dydek2011}. \textcolor{black}{For the lowest salt concentration ($10^{-4}$ M), a non-constant over-limiting conductance could also be related to additional ion transport from the dissociation of water or dissolved CO$_2$ in the solution, but transient diffusion is the more likely cause of the observed weak oscillations around the mean linear profile of OLC at all salt concentrations, as discussed above and in the Supporting Information.}

\subsection{ Limiting current }

To the best of our knowledge, this is the first quantitative fit of experimental data by a theory of OLC, by any mechanism.   In order to check our assumptions and identify the physical mechanism for OLC,  the scalings of $I_{lim}$, $R_s$, and $\sigma_{OLC}$ are investigated with respect to concentration of CuSO$_4$ over four orders of magnitude, from 0.1 mM to 1.0 M.
{
The error in each data point (from the Supporting Information) is of order 10\%, which is smaller than the point size in the log-log plots of \textcolor{black}{\ref{fig:3}d-f} showing power-law scalings.
}

According to dilute solution theory, \ref{eq:I_lim}, the limiting current, $I_{lim}$, is linearly proportional to concentration. The fitted $I_{lim}$ (\ref{fig:3}d) verifies this scaling at low concentration and deviates to lower values at high concentration, consistent with reduced Cu$^{2+}$ activity and diffusivity~\cite{quickenden1996}. A simple estimate (red curve), using free-solution values~\cite{noulty1987} $D_0$ for $D(c_0)$ in \ref{eq:I_lim}, captures the scaling of the data for $I_{lim}(c_0)$ well. In the Bruggeman approximation, however, the macroscopic  diffusivity $D_m=\epsilon^{3/2} D_0$ at low concentration is $\approx 13$  times smaller than the apparent dispersion $D$ from ~\ref{eq:I_lim}.  The discrepancy partly reflects  transient diffusion (effectively smaller $L$) since the diffusion distance $2\sqrt{D_m t}\approx 0.5$ mm when limiting current is reached  is somewhat smaller than the frit thickness, $L=1$ mm.

Consistent with our analysis of OLC below, the leading cause of the enhanced dispersion inferred from $I_{lim}$ may be electro-osmotic convection in the glass frit.  Electro-osmotic flow toward the impermeable membrane/cathode structure is balanced by a pressure-driven back flow that produces dispersion~\cite{yaroshchuk2011}. Taylor dispersion is negligible based on the formula for a single cylindrical pore, $D_{Taylor}/D_m-1=\epsilon_p^{3/2}\mbox{Pe}_1^2/48\sim  0.01$, even using a large EOF velocity $U=400 \mu$ $m/s$ in the single-channel P{\' e}clet number $\mbox{Pe}_1= Uh_p/D_0$. For a network of pores, however, there is additional dispersion due to randomness in streamline topology~\cite{Koch1985} (also referred to as ``eddy diffusion" or dispersion~\cite{vanDeemter1956}). Indeed, the simple estimate $D_{eddy}/D_m\approx\mbox{Pe}_e/2\approx 10$ could explain the fitting result for $D$ above where, we use the same velocity and a mean loop size $h_{loop}=25$ $\mu m$ consistent with \ref{fig:2}c in estimating the eddy P{\' e}clet number, $\mbox{Pe}_e = U h_{loop}/D_m$.
We have also confirmed that replacing the silica glass frit with a loopless porous medium (an anodic aluminum oxide membrane with straight pores of 200 nm diameter) leads to a reduced $I_{lim}$ consistent with $D_m \approx D_0$, \textcolor{black}{as will be reported elsewhere~\cite{jihyung_preprint}. }

\subsection{ Series resistance }

The series resistance $R_s$ can be attributed to two primary sources: the Ohmic resistance of the reservoir, $R_{res}$ at low salt, and the Faradaic resistance of the anode, $R_F$, at high salt. (The cathode Faradaic resistance is reduced by contact with the Nafion membrane.) Neglecting concentration polarization and assuming equal ionic diffusivities, we estimate the reservoir resistance using
\begin{equation}
R_{res} = \frac{k_BT \textcolor{black}{L}}{ 2 \pi R^2 (ze)^2 D(c_0) c_0  }.
\end{equation}
Assuming linearized Butler-Volmer kinetics~\cite{bard_book}, we estimate the anode Faradaic resistance as
\begin{equation}
R_F = \frac{k_BT}{ne I_0} = \frac{k_BT}{2e K_0 c_0^{1/2}}
\end{equation}
where $I_0 = K_0 c_0^{\alpha_c}$ is the exchange current density. The transfer coefficient is $\alpha_c=\frac{1}{2}$ for the rate-limiting transfer of one out of $n=2$ electrons~\cite{mattsson1959,brown1965}.  The data in \ref{fig:3}e is quantitatively consistent with the theory, $R_s = R_{res}+R_F$ (red curve), with a fitted prefactor of $K_0=2.8$ A/m$^2$ (for Molar $c_0$). The measured exchange current $I_0=3.9$ A/m$^2$ at $c_0=0.5$M is close to $I_0=9.7$ A/m$^2$ from recent experiments with CuSO$_4$ at neutral pH~\cite{semichaevsky2010} (which is below $I_0$ in strong acids~\cite{mattsson1959,brown1965}).  We are thus able to attribute the remaining voltage $V$ in \ref{eq:Vcorr}, which exhibits OLC (\ref{eq:addsurface}), to the glass frit.

\subsection{ Overlimiting current }

The SC and EOF mechanisms for OLC are distinguished by different scalings in \ref{eq:scalings} with salt concentration $c_0$ and surface charge density, $q_s$. In the absence of flow, the over-limiting conductance from surface conduction is given by \ref{eq:SC}~\cite{dydek2011}.  {\color{black} Using $D=\epsilon_p^{3/2} D_m$,
the theoretical line with constant $q_s$ (green dashed line) roughly matches the experimental value at the lowest salt concentration, but lies far below the data at higher concentrations (black points)  in \ref{fig:3}f.  This is consistent with the prediction that SC becomes important only at low salt concentrations~\cite{dydek2011}, but we must also consider the effect of charge regulation.
}
It is well known that that the surface charge of silica is regulated by the dissociation of silanol groups~\cite{behrens2001,heyden2005,gentil2006},
\begin{equation}
\mbox{SiOH} \longleftrightarrow \mbox{SiO}^- + \mbox{H}^+ \ \  \mbox{(pK=7.5)}
\end{equation}
Using the Gouy-Chapman-Stern model to obtain the surface pH~\cite{behrens2001}, the surface charge can be calculated from the measured bulk pH versus salt concentration,
assuming pK=7.5 for silica.
 \textcolor{black}{(Details are in the supporting information, where it is also shown that pK has little effect on the prediction of OLC.)}
}%
As shown in \ref{fig:3}f (green dot-curve), the decrease of $q_s$ with increasing $c_0$ leads to the opposite trend from the experimental data. We conclude that SC is not the primary mechanism for OLC in our experiments, although it may contribute at low salt concentration.

{\color{black}
Since the data cannot be explained by SC, the theory suggests that EOF is the likely mechanism.
In a parallel-plate microchannel, the transition from SC to EOF is predicted to occur at a thickness $8 \mu$m (for $c_0=1$M, $L=1$mm)~\cite{dydek2011}, which is comparable to the mean effective cylindrical pore diameter of the silica frit, $4h_p=2.3 \mu$m. As noted above, the single-channel analysis under-estimates the true effect of EOF in a heterogeneous porous medium since there are eddies around loops in the pore network (\ref{fig:EOF}), whose size $h_e\gg  h_p$ could reach tens of microns in our silica frit, thus making EOF dominant.  

Indeed, the EOF scaling theory is consistent with the experimental data.  
If we fit $h_e$ to the lowest concentration data point using the theoretical formula \ref{eq:EOF} with $D_m=(0.4)^{3/2} D_0$ and $D_0=8.5 \times 10^{-6}$ cm$^2/$s~\cite{noulty1987} and all other parameters known, we obtain an eddy size of $h_e=100 \mu$m, comparable to the thickness of the depleted region~\cite{dydek2011}, $(1-I_{lim}/I)L$ for $I = 1.1 I_{lim}$, but this neglects the unknown numerical prefactor in \ref{eq:EOF}.  We expect this prefactor to be smaller than 1 in order to obtain a mean eddy size in the 5-50 $\mu$m range, consistent with the scale of loops in the glass frit (set by aggregates of sintered particles seen in the SEM images (\ref{fig:2})). This range is also consistent with the characteristic eddy size inferred from effects of hydrodynamic dispersion on $I_{lim}$ above.

Remarkably, without any adjustable parameters, the EOF scaling theory \ref{eq:EOF} accurately predicts the observed dependence on salt concentration, varying over four orders of magnitude, $c_0 = 0.1$ mM - 1.0 M. 
}
With constant $q_s$ (red dashed line in \ref{fig:3}f), the experiments reveal the nontrivial scaling, $\sigma_{OLC} \sim c_0^{4/5}$,  at low salt concentration, and the predicted effect of  surface charge regulation $\sigma_{OLC}\sim q_s(c_0)^{2/5}$ also captures the trend at high concentration (red solid curve in \ref{fig:3}f).
We conclude that EOF is the likely mechanism for OLC in our experiments.

{\color{black}

\subsection{ Surface charge modification}

The experimental data are consistent with the new hypothesis of OLC by double-layer-driven transport in a porous medium. The key property enabling OLC is the surface charge of the pores, while the geometry plays a secondary role in selecting the physical mechanism as SC, EOF or some combination of both. In the preceding experiments, the surface charge is not controlled, but its value is determined from the measured pH through the well-studied charge regulation of silica in water through proton adsorption reactions~\cite{behrens2001,heyden2005,gentil2006}. The resulting trend in surface charge agrees well with the EOF scaling theory, but we also report definitive proof of the effect of surface charge by two different means of surface chemical modification. 

First, we alter the surface charge by silanization to obtain a stable positive charge~\cite{jani2009,jani2010}. The glass frit is hydroxylated in boiled hydrogen peroxide (30$\%$ H$_2$O$_2$) for 30 min. After rinsing in distilled water, the membrane is dried under nitrogen gas, and then the membrane is immersed in toluene (Aldrich) containing 1.5$\%$ of (3-Aminopropyl)triethoxysilane (APTES) (Aldrich) solution for 2 hours under nitrogen atmosphere at room temperature. Next, the silanized membrane is thoroughly rinsed with toluene, ethanol, and distilled water and dried under a stream of nitrogen, and the APTES modified membrane is cured overnight in an oven at 120 $^{o}$C.

Second, we perform layer-by-layer (LbL) deposition of charged polymer monolayers~\cite{Ai2003,Yeo2012}. The glass frit is treated under oxygen plasma for 5 minutes to generate negative charge. The negatively charged glass frit is then immersed in polycation solution, 1 mg/mL polyallylamine hydrochloride (PAH) in 500 mM NaCl at pH 4.3, for 30 min to produce a stable positive surface charge. The PAH modified membrane is thoroughly rinsed with distilled water three times (for 10 min. each).

\begin{figure}
\includegraphics[width=3.7in]{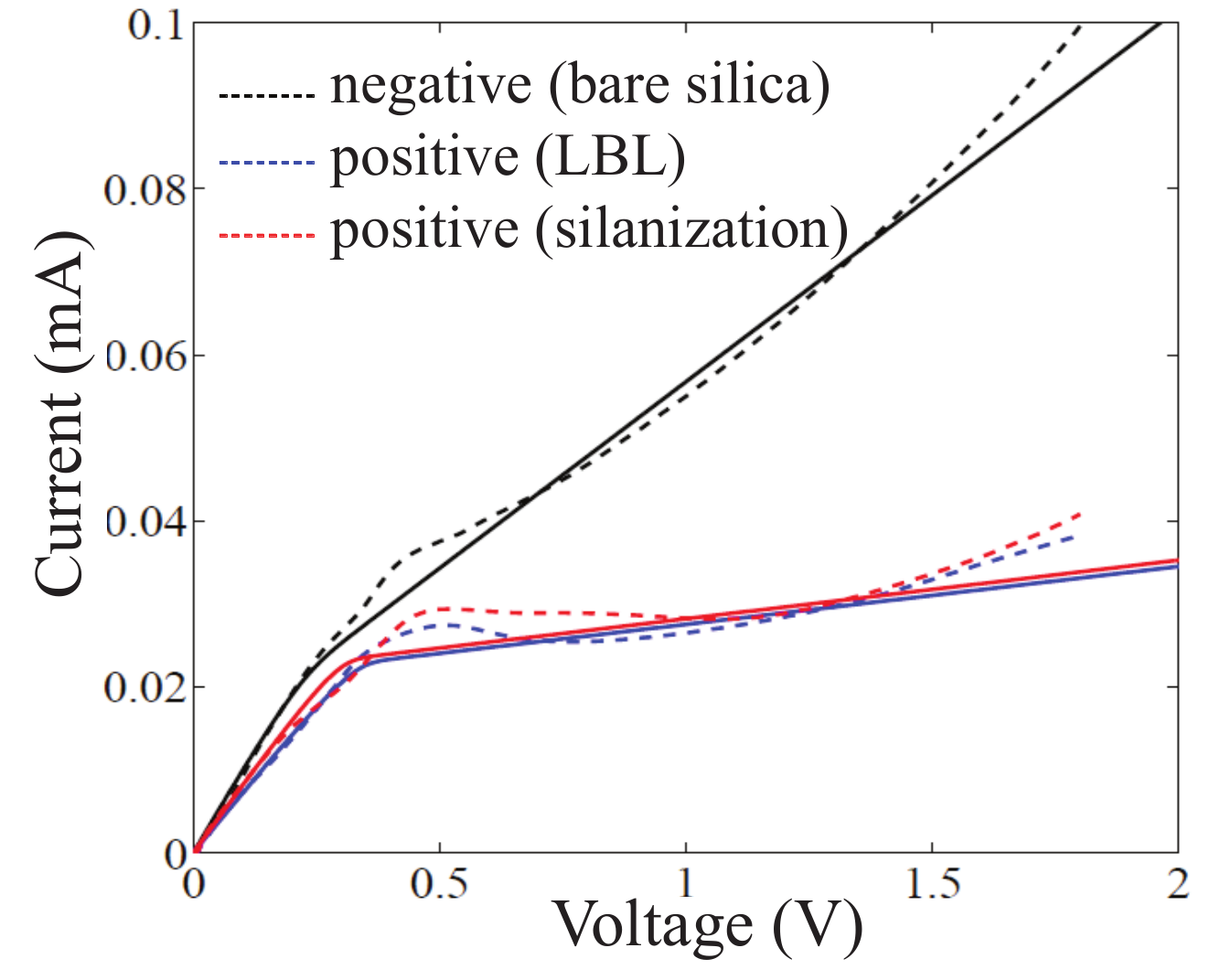}
\caption{ \textcolor{black}{Effect of surface charge modification.  The current-voltage relations obtained by linear sweep voltammetry at 2mV/s for 1 mM aqueous CuSO$_4$ solution in the same apparatus above are shown for negatively charged  bare silica compared with two positively charged surface layers, obtained by silanization and layer-by-layer PAH polymer deposition. Dashed curves indicate the data, and solid lines the fit to the theoretical steady current-voltage relation, \ref{eq:addsurface}.  }}
\label{fig:surface}
\end{figure}

The effect of these surface modifications on the current-voltage response of the glass frit are shown in \ref{fig:surface}.   Again, each individual curve fits the same formula, \ref{eq:addsurface} and \ref{eq:Vcorr} with three parameters ($I_{lim}$, $R_s$, and $\sigma_{OLC}$). As $I_{lim}$ is associated with bulk transport and $R_s$ with reservoir and electrode resistances, both are expected to be roughly independent on the surface charge density, although surface conduction at low voltage (linear response) leads to a slightly increased resistance for positive versus negative charge, as observed. The crucial parameter $\sigma_{OLC}$ is directly related to surface charge and is largest for negative charge, according to the theory.  This is consistent with the observation of reduced OLC by one order of magnitude from negative to positive surface charge.  

An unexpected finding is that the positive charge-modified frits exhibit nonzero OLC, albeit much smaller than that of the negative bare silica frit.  For SC in nanopores, this would not be possible, as the surface conductivity would have the wrong sign to sustain OLC after bulk salt depletion. For EOF in larger pores, in retrospect, it is clear that OLC is still possible with positive charge because the eddies of electro-osmotic convection can still transport some salt faster than diffusion to the membrane, only with the opposite sense of rotation and while fighting against the surface conduction. This shows the complexity of OLC in porous media in the EOF regime and the need for new models.  
}

\begin{figure}
\includegraphics[width=4in]{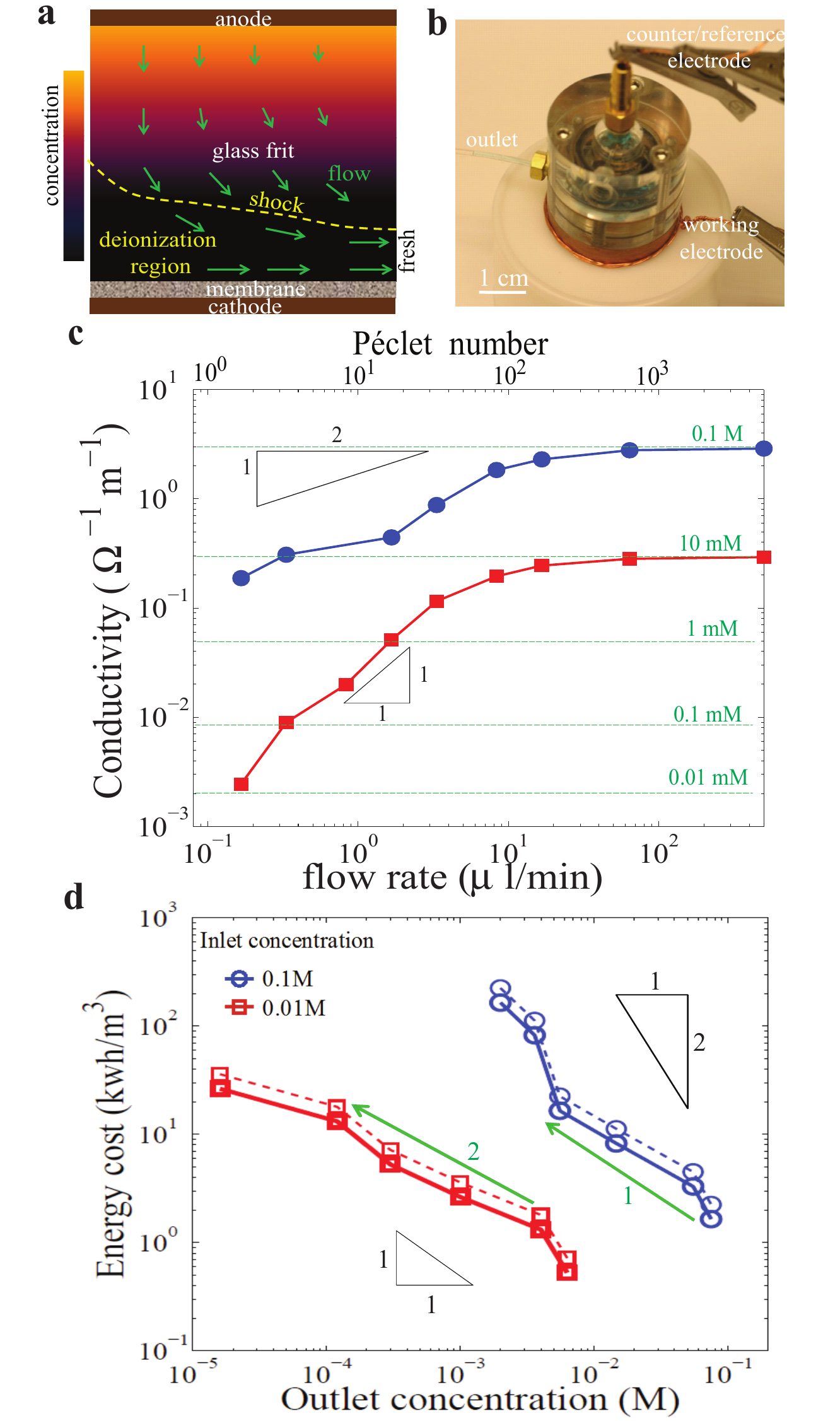}
\caption{Water deionization by shock electrodialysis operating under OLC. (a) Schematic of the extraction flow and salt concentration profile. (b) Photograph of the button-cell device. (c) Conductivity (y-axis) and salt concentration (green lines) of the extracted water versus flow rate, with theoretical scalings from \ref{eq:f}. (d) Energy cost per volume with (dash lines) and without (solid lines) the series resistance attributed to the reservoir and electrodes for the same experiments. }
\label{fig:4}
\end{figure}

\subsection{ Water deionization }

To complement the electrochemical evidence, we experimentally verify the extreme salt depletion associated with OLC~\cite{dydek2011} by driving a flow that produces a deionization shock~\cite{mani2011}. In microfluidic devices with tiny (nL) volumes, salt concentration is typically inferred by optical detection of fluorescent particles~\cite{wang2005,kim2007,yossifon2008,kim2010,zangle2009}, but here we can directly extract  macroscopic (0.01-1.0 mL) samples of deionized water from the glass frit  (\ref{fig:4}a) and test their conductivity by impedance spectroscopy. A proof-of-concept device (\ref{fig:4}b) is equipped with
{
a circular outlet ($d=0.5$ mm in diameter)
at one point on the side of the frit just above the membrane,
leading to an annular collection channel connected to the device outlet (\ref{fig:2}e).
}%
The volumetric flow rate ($Q$) is controlled by a syringe pump.  Two sets of experiments are conducted with initial concentrations of $100$ mM and $10$ mM, holding the applied voltage constant at $V_{app}=1.5$ V, well into the OLC regime (\ref{fig:2}).
{
The current remains steady for hours, indicating stable continuous operation during deionization. (See the Supporting Information on chronoamperometry.)
}%

At low flow rates, we find that the salt concentration can be reduced by four orders of magnitude to 15 $\mu$M (\ref{fig:4}c).  Essentially all of the CuSO$_4$  ions are removed, down to the level of the water ions (pH$\sim 5.5$) and below the U.S. regulatory limit for copper in drinking water ($<0.02$ mM) \cite{USEPA2007}.  As in \ref{fig:1}d, the region of deionization ($>0.5$ mm) near the outlet extends across more than half of the frit thickness ($1.0$ mm) and is maintained in the outgoing flow. This establishes the existence of a stable deionization shock propagating against the flow (in the moving frame of reference)~\cite{mani2009,zangle2009} over a macroscopic distance in the porous medium~\cite{mani2011}.
{
Such extreme deionization propagating so far into the frit cannot be explained by theories of ED based on convection-diffusion in neutral electrolytes~\cite{sonin1968,probstein1994}.
}%

{
This observation suggests the possibility of harnessing deionization shocks in porous media for water purification.  Although our apparatus has not been optimized for this purpose, it serves to illustrate the principles of ``shock electrodialysis".  The basic idea is to drive over-limiting current through a porous medium and extract deionized water between the membrane and the shock with a pressure-driven cross flow.  In a scalable system for continuous operation (discussed below),  additional outlets must also collect brine from the frit (deflected by the shock) and reaction products from electrode streams (such as hydrogen and oxygen from water splitting, as in standard ED).  Here, we have just one, small fresh-water outlet and negligible brine accumulation at the anode, but this
}%
suffices to demonstrate the general tradeoff between flow rate and deionization~\cite{dydek_thesis,mani_crossflow} (\ref{fig:4}c): For a given geometry and current, the flow rate must be small enough to allow the shock to propagate across the outlet, in order to fully deionize the outgoing stream.  As the flow rate is increased, the shock retracts toward the membrane and crosses the outlet, thereby causing the salty fluid from the diffusion layer to be mixed with the deionized fluid behind the shock.

\subsection{ Flow rate dependence }

At fixed voltage, the deionization factor $f=c_0/c_{out}$ (ratio of inlet to outlet salt concentrations) is controlled by the P{\' e}clet number,
$\mbox{Pe} = \frac{ U d } {D} =  \frac{ Q }{ d D},$
where $U$ is the mean outlet velocity.  In our apparatus (\ref{fig:2}e), asymmetric flow leads to complicated concentration profiles (\ref{fig:4}a), but we can use similarity solutions for simple uniform flows to understand the scaling of $f$ for $\mbox{Pe}\gg 1$ (\ref{fig:4}c).  For the SC mechanism, the  shock has a self-similar nested boundary layer structure consisting of an outer convection-diffusion layer (or diffusive wave~\cite{bazant1995}) and an inner depleted region, whose overall thickness (distance from the membrane) scales as $\mbox{Pe}^{-\gamma}$, where $\gamma=1$ for uniform normal flow through the membrane~\cite{dydek_thesis,mani2011} and $\gamma=\frac{1}{2}$ for uniform cross flow along the membrane~\cite{mani_crossflow}.  Integrating the self-similar concentration profile over a fixed-diameter outlet then implies the scaling
\begin{equation}
f = \frac{c_0}{c_{out}} \sim \mbox{Pe}^{-\gamma} \sim \left( \frac{ D} {Q} \right)^\gamma  \label{eq:f}
\end{equation}
with $\frac{1}{2}\leq \gamma \leq 1$ for the SC mechanism with pressure-driven flow.  For the EOF mechanism in a microchannel without net flow~\cite{dydek2011}, the depleted region has nearly uniform mean concentration scaling as, $c_d/c_0 \sim (I/I_{lim})^{\nu}$, where $\nu\approx 0.3-0.4$.  Although the theory needs to be extended for porous media and pressure-driven net flow, this result with $I/I_{lim}\sim\mbox{Pe}^{\gamma}$ (for the convection-diffusion layer) and $f\sim \tilde{c}_d^{-1}$ (if the depleted region spans the outlet) suggests that the exponent $\gamma$ in \ref{eq:f} may be replaced by the smaller value $\gamma\nu$ for EOF.

Both sets of experiments show the expected trend of the deionization factor with the flow rate (\ref{fig:4}c). The conductivity of the inlet and outlet solutions is measured by impedance and calibrated against solutions of known salt concentration. (See the Supporting Information.) At the lowest flow rate, of order $0.1\mu$L/min, we obtain $f>10$ starting from $c_0=0.1$ M and $f\approx 10^{2}$ starting from $c_0=10$ mM.  At each flow rate, the solution with the lower initial ion concentration ($10$ mM) consistently yields greater percentage reduction of conductivity (or concentration) than that of the higher initial ion concentration ($100$ mM). The larger deionization factor results from the larger dimensionless current ($I/I_{lim}$) and more extended deionization region (or shock) at lower salt concentrations, consistent with the theory~\cite{dydek2011}.  This trend is also a consequence of mass balance, $f \sim I/(Q c_0)$, as in standard ED.

\subsection{ Energy efficiency }

The energy cost per volume of deionized water in the experiments of \ref{fig:4}c is plotted in \ref{fig:4}d versus the outlet concentration $c_{out}$. Comparing the energy cost with (dashed line) and without (solid line) the electrode and reservoir series resistances shows that less than half of the total energy cost is spent driving the copper reactions.  As in standard ED, such electrode resistances can be made negligible compared to a larger total voltage in a scalable, multilayer system (\ref{fig:stack}).   As indicated by the green arrows, the button cell can desalinate brackish water ($0.1$ $M$) to produce potable water ($<10$ mM) at a cost of $\approx 10$ kWh/m$^3$  and then deionize close to $0.01$ mM in a second step at roughly the same cost. The net energy cost of $\approx 20$ kWh/m$^3$ is well above the thermodynamic limit of $\approx 0.15$ kWh/m$^3$, but this is mainly a consequence of the experimental geometry, which was not designed for this purpose.

To boost the efficiency in a practical shock ED system, the cross flow must cover as much of the active area (drawing current) as possible.
Since our device has a point outlet from the frit at only one azimuthal angle, rather than gap spanning its circumference,
fluid is only extracted from a very small area $\approx \pi d^2/2$, roughly $1/50$ of the total cell area $\pi R^2$.  As a result, the total power use is nearly independent of the flow rate, and the energy/volume $=$ power/(flow rate) should scale as $\mbox{Pe}^{-1} \sim f^{1/\gamma}$ from \ref{eq:f}, which consistent with the data in \ref{fig:4}d.  With uniform cross flow covering the entire active area as in \ref{fig:stack}, the energy cost could, in principle, be reduced by the same factor to $\approx 1$ kWh/m$^3$.  This suggests that shock ED
has the potential to be competitive with other approaches on efficiency, while having some other possible advantages in separations discussed below.

\begin{figure}
\includegraphics[width=3in]{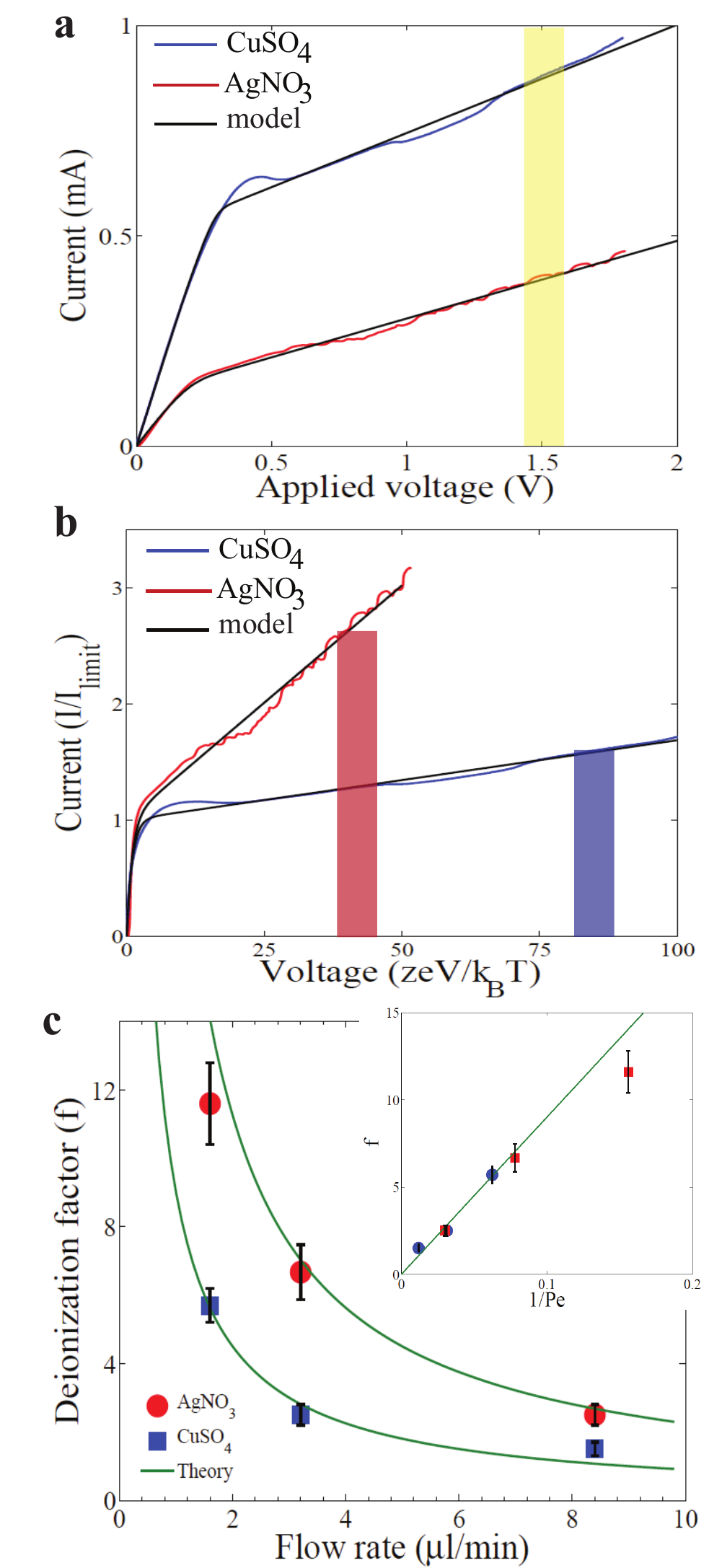}
\caption{  Overlimiting current and shock electrodialysis with $10$ mM silver nitrate using the device of \ref{fig:2}.
The current-voltage relation in (a) exhibits a similar, constant over-limiting conductance, similar to $10$ mM copper sulfate, but with smaller limiting current and voltage scale, which lead to a larger dimensionless current  in (b).  Bars indicate data taken at $V_{app} = 1.5$ V.  The deionization factor in (c) is larger in AgNO$_3$ versus CuSO$_4$ at the same voltage, flow rate and salt concentration, as predicted by the scaling theory, \ref{eq:f} (solid curves and inset data collapse).  }
\label{fig:Ag}
\end{figure}

\subsection{ Electrolyte dependence } 

Until this point, our copper electrolytic cell has provided a convenient model system to establish the basic principles of shock ED, but the method is much more general and can be applied to arbitrary electrolytic solutions.  As in standard ED, the electrodes can be chosen to drive any desired brine-producing reactions, such as water electrolysis, while the current is carried across a stack of many membranes by the input solution (see below).
In our device, we have only one separation layer and copper electrodes, but we can consider a different electrodeposition reaction. In the same device, we remove silver ions Ag$^+$ from 10 mM silver nitrate (AgNO$_3$) through the porous frit and Nafion membrane by silver deposition at the cathode. The current is sustained by copper dissolution at the anode without allowing sufficient time for concentration polarization to cross the reservoir and reach the frit during a voltage sweep.

The results in \ref{fig:Ag} are similar for both electrolytes and consistent with the theory, thereby showing the generality of the phenomenon. The raw current-voltage data (\ref{fig:Ag}a) indicates a slightly smaller over-limiting conductance and much smaller limiting current for $AgNO_3$, as suggested by the scaling $I_{lim}\sim z D$.  ($D_{\mbox{AgNO}_3}=1.68\times10^{-5} \mbox{cm}^2/s$ ~\cite{albright1972} and  $D_{\mbox{CuSO}_4}=6.75\times10^{-6} \mbox{cm}^2/s$
~\cite{noulty1987} at $10$ mM.) At the same voltage $V_{app}=1.5$ V (bars in \ref{fig:Ag}b), the dimensionless voltage $\tilde{V}$ is also smaller by a factor of two (for monovalent versus divalent cations), and the dimensionless current $\tilde{I}=2.6$ for AgNO$_3$  is larger than $\tilde{I}=1.6$ for CuSO$_4$, which implies a wider depletion zone, scaling as $(1-\tilde{I}^{-1})$. During water extraction, the dependencies of the deionization factor on flow rate and diffusivity are nicely captured by the simple scaling of \ref{eq:f}, as shown in \ref{fig:Ag}c.

\section{Discussion }

Our primary finding is that thin double layers in porous media can enable faster ionic transport, leading to new surface-driven mechanisms for OLC based on SC and EOF. In particular, electro-convection driven by EOF can sustain OLC in a heterogeneous porous medium with micron-scale pores pressed against an impermeable electrodialysis membrane.   The onset of OLC is associated with a macroscopic region of deionization within the pores (outside the double layers), which can propagate against a pressure-driven flow like a shock wave. In steady state, the  over-limiting conductance is approximately constant (aside from surface charge regulation), in spite of enormous spatial variations in conductivity (up to three orders of magnitude). These surprising phenomena are in stark contrast to the constant conductivity of ion-exchange membranes with smaller pores and overlapping (thick) double layers.

\begin{figure}
\includegraphics[width=0.65\columnwidth,keepaspectratio=true]{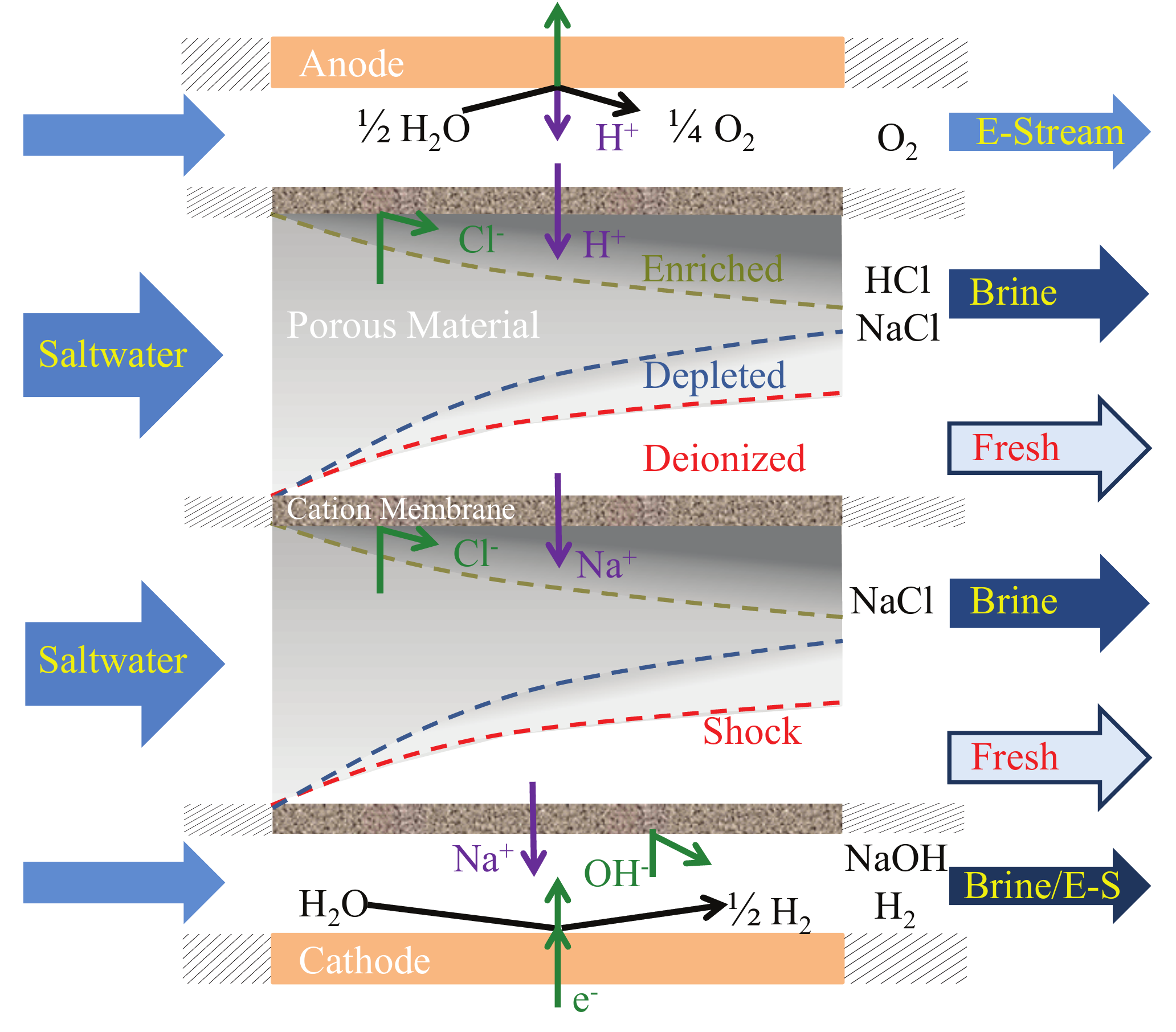}
\caption{
Sketch of a scalable shock electrodialysis system, motivated by our experimental results.
\label{fig:stack} }
\end{figure}

The nonlinear electrokinetic properties of porous media can be exploited for separations.  Our proof-of-concept experiments (\ref{fig:4}, \ref{fig:Ag}) show that deionized water can be continuously extracted from salty water via a porous medium sustaining OLC.  \textcolor{black}{It is beyond the scope of this paper to build and test a practical desalination system with electrode streams, but }
a possible  design for a scalable shock ED system is shown in \ref{fig:stack}. A stack of two (or more) separation layers of negatively charged porous media separated by cation exchange membranes sustains an over-limiting current. In each layer, the input solution (e.g. NaCl) undergoes salt enrichment near one membrane and salt depletion with a deionization shock near the other. In a pressure-driven cross flow, these regions are continuously separated into fresh and brine streams upon leaving the porous medium. By varying the position of the fresh/brine stream splitting in each porous layer, high water recovery (wide shock) can be traded against low energy cost (thin shock).
As in standard ED, direct current can be sustained at the electrodes by water splitting (or other) reactions, whose overpotential becomes negligible compared to the total voltage as the stack size increases.

Besides water deionization, such a system may also find applications in brine concentration (e.g. for salt precipitation or forward osmosis) or in nanoparticle separations. Since the separation occurs within the porous medium in cross flow, the membrane removing ions is electrokinetically shielded and may resist fouling (which is a concern for other desalination methods~\cite{potts1981,shannon2008}). Membraneless designs with layered porous media of different pore sizes (analogous to micro/nano channel junctions~\cite{kim2007,kim2010,zangle2009}) may also be possible. Clogging by incoming particles or brine precipitates could be managed by reverse bias, cleaning, or replacement due to the low cost of the porous materials themselves.

By combining microfiltration and deionization in one step, shock ED may also enable more compact, portable, point-of-use systems. Besides filtration by size, suspended particles are also strongly filtered by charge. Co-ionic particles (with the same charge as the pore walls and the membrane) are repelled by the shock~\cite{wang2005}, but counter-ionic particles are accelerated through the depleted region by the large electric field and sent to the outlet, if they are blocked by the membrane.  Some of these advantages are also possessed by microfluidic desalination devices with aligned flow and current in individual microchannels~\cite{kim2010}, but with higher fabrication costs and smaller flow rates (even with massive parallelization).  By decoupling the flow and current directions using porous media, it is possible to cheaply extend deionization and filtering over macroscopic volumes.

Selective ion exchange and separation may also be possible by shock ED. In contrast to existing methods for heavy metal removal based on adsorption in nanocrystals \cite{Yavuz2006}, functionalized porous media \cite{Davis2002,Feng1997}, and biosorbents \cite{volesky1995,WanNgah2008},
shock ED is not limited to particular ions and exhibits ion selectivity (based on surface transport in porous media), which could be used to fractionate different metal ions  and/or charged macromolecules by splitting streams in cross flow through the cell.   Multivalent/monovalent ion separation can also be achieved
by electro-osmotic convection in nanochannels~\cite{pennathur2005a,pennathur2005b} or
by capacitive charging of porous electrodes~\cite{zhao2012}, but  shock ED could enable continuous, scalable separations based on both size and charge.

\section*{Acknowledgements}
This work was supported by a grant from Weatherford International through the MIT Energy Initiative. B. Z. and M. Z. B. also acknowledge support from the USAÐIsrael Binational Science Foundation (grant 2010199) and J.-H. H.  support from the Basic Science Research Program through the National Research Foundation of Korea (NRF) funded by the Ministry of Education (2012R1A6A3A03039224). The authors thank P. Morley and A. Gallant at the MIT Central Machine Shop for producing the prototype.  

\section*{Author contributions}
All authors contributed to the research. D. D. led the experimental effort and analyzed the data. M. Z. B. suggested the approach and led the theoretical interpretation and writing.

\section*{Additional information}

The authors declare no competing financial interests.

\bibliography{elec38_langmuir}

\clearpage

\section*{ Supporting Information }

\subsection*{1. Sweep rate effect for linear sweep voltammetry}
It is not possible to make a steady-state measurement in our system because the dominant concentration polarization of the reservoir (salt depletion near the frit and enrichment near the anode) is very slow.  As a reasonable and reproducible alternative, we managed to obtain quasi-steady current-voltage curves by linear-sweep voltammetry. In general, the curves depend on the voltage sweep rate, but it is possible to select a sweep rate that is small enough such that quasi-steady-state conditions are reached in the frit, but large enough to avoid significant reservoir concentration polarization. 

\begin{figure}[h]
\includegraphics[width=0.5\columnwidth,keepaspectratio=true]{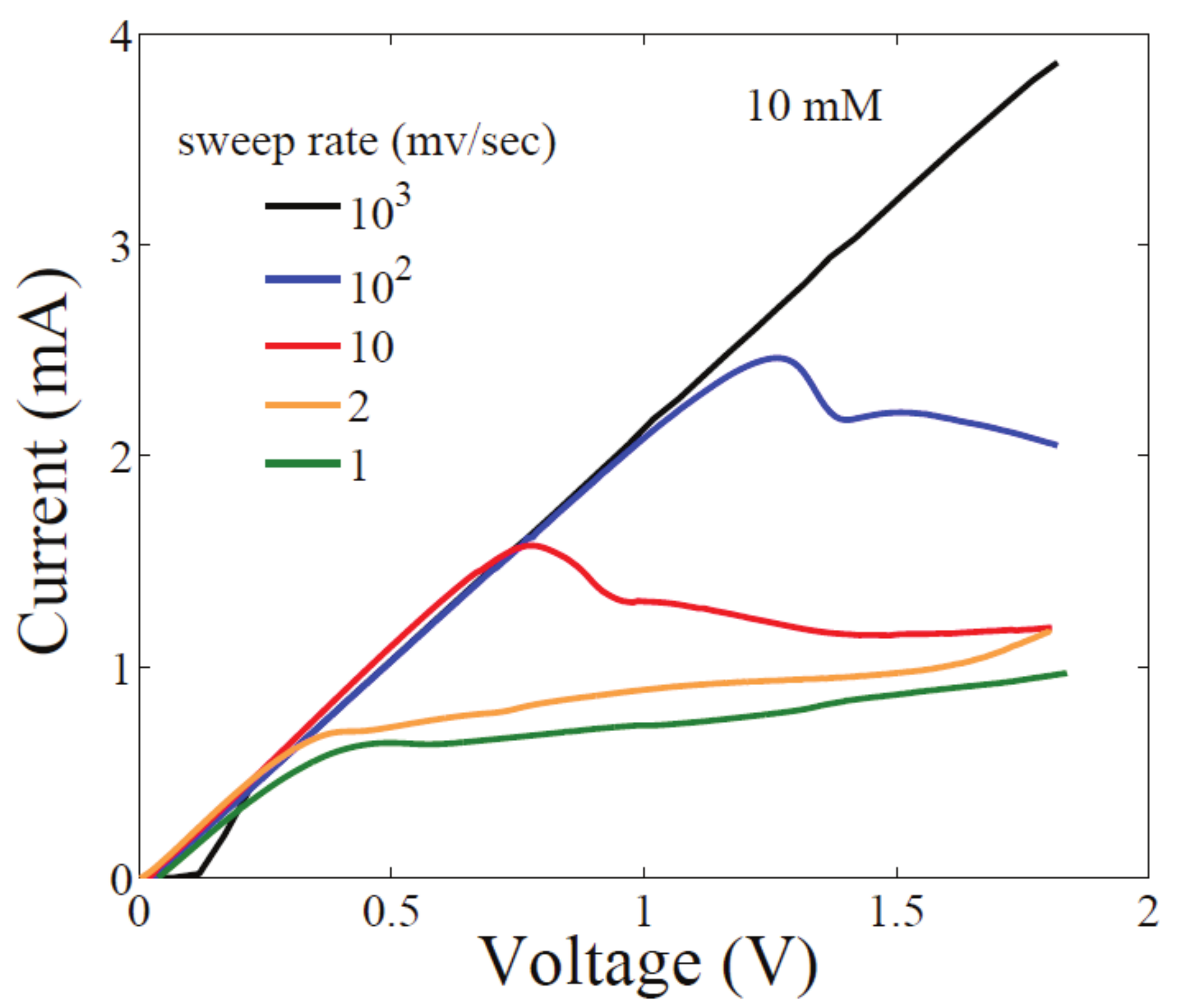}
\caption{Sweep rate effect for measured current-voltage curves using linear sweep voltaetry in a $10$  mM aqueous CuSO$_4$ solution. The optimal sweep rate is about $1$ or $2$ mV/sec.}
\label{fig:SFsweeprate}
\end{figure}

A sample of our current-voltage data, with variable sweep rate, is shown in Fig. \ref{fig:SFsweeprate} for a 10 mM CuSO$_4$ solution.  The data indicates that a sweep rate of approximately 2 mV/sec marks the transition between high and low sweep rate regimes. This sweep rate was used to obtain all the data reported in Fig. 5 of the main text. The time to reach overlimiting current corresponds approximately to the theoretical diffusion time across the frit.  Above this sweep rate, the concentration becomes depleted inside the frit near the membrane, leading to a sudden bump in voltage and departure from Ohmic behavior. Below this sweep rate, the under-limiting portion of the curve begins to deviate from classic Ohmic linear dependence due to concentration polarization in the reservoir.

\subsection*{2. Reproducibility of the measured current-voltage curves}
\begin{figure}[h]
\includegraphics[width=\columnwidth,keepaspectratio=true]{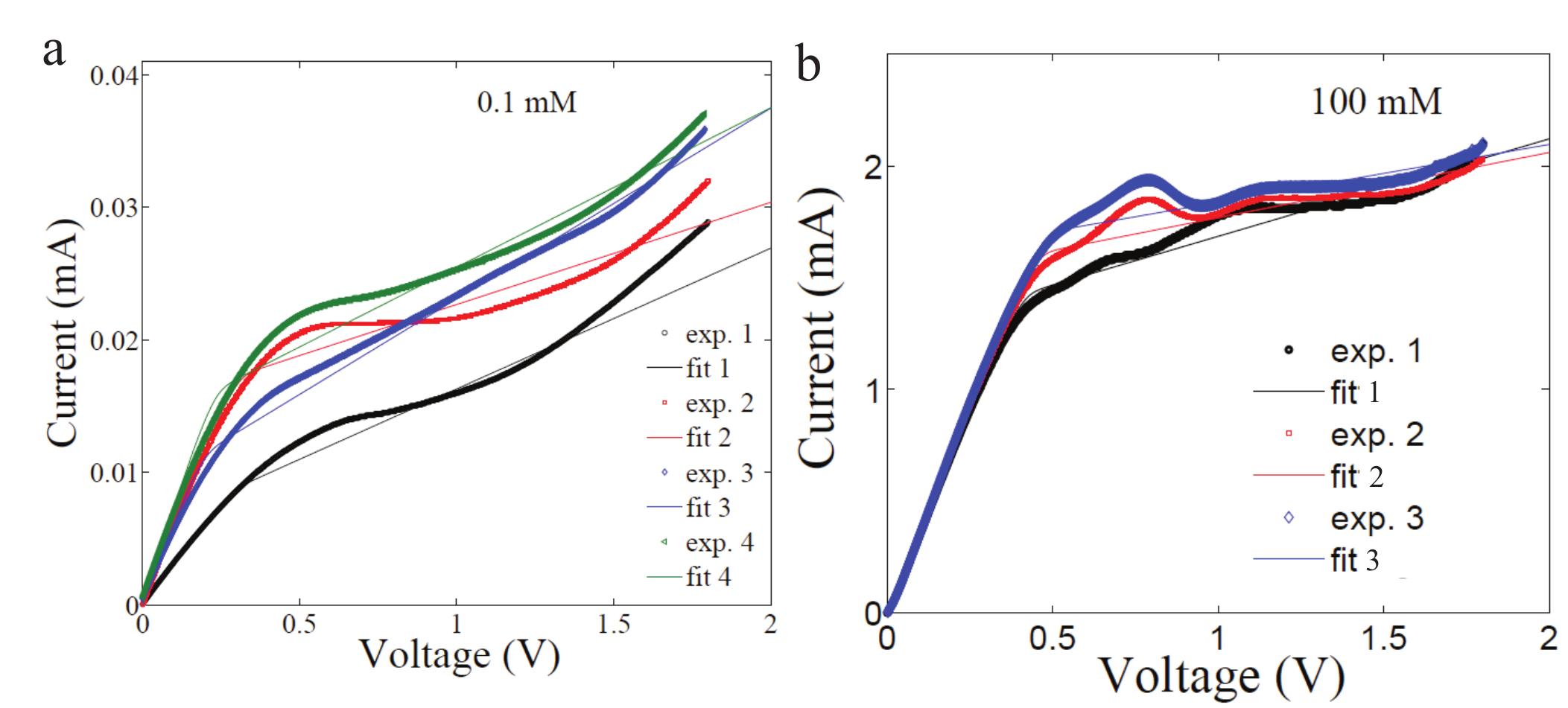}
\caption{Reproducibility of current-voltage curves at the fixed rate for (a) $0.1$ mM CuSO$_4$ solution and (b) $100$ mM CuSO$_4$ solution.}
\label{fig:SFreproducibility}
\end{figure}

The results are reproducible.  For example, we repeated the measurement of current-voltage curves at a fixed sweep rate four times with a $0.1$ mM CuSO$_4$ solution and three times with a $100$ mM CuSO$_4$ solution, as shown in Fig. \ref{fig:SFreproducibility}. The three parameters -- limiting current $I_{lim}$, series resistance $R_s$, and over-limiting conductance $\sigma_{OLC}$ -- were obtained by fitting the theory of the quasi-steady current-voltage response to the measured curves. In spite of some systematic oscillations in the data, associated with the slow voltage sweep (see above), the relative error of these three parameters for a $100$ mM CuSO$_4$ solution is $10\%$, $1\%$, and $30\%$, for limiting current, series resistance, and over-limiting conductance, respectively. The relative error  of these three parameters for a $0.1$ mM CuSO$_4$ solution is $30\%$, $40\%$, and $28\%$, for limiting current, series resistance, and over-limiting conductance, respectively. These uncertainties are much smaller than the order-of-magnitude variations in parameters calculated from experiments with different salt concentrations. (In particular, the errors are smaller than the point sizes used in Figs. 5 and 7 in the  main text.)

\subsection*{3. Comparison of current-voltage curves with and without a glass frit}
\begin{figure}[h]
\includegraphics[width=0.5\columnwidth,keepaspectratio=true]{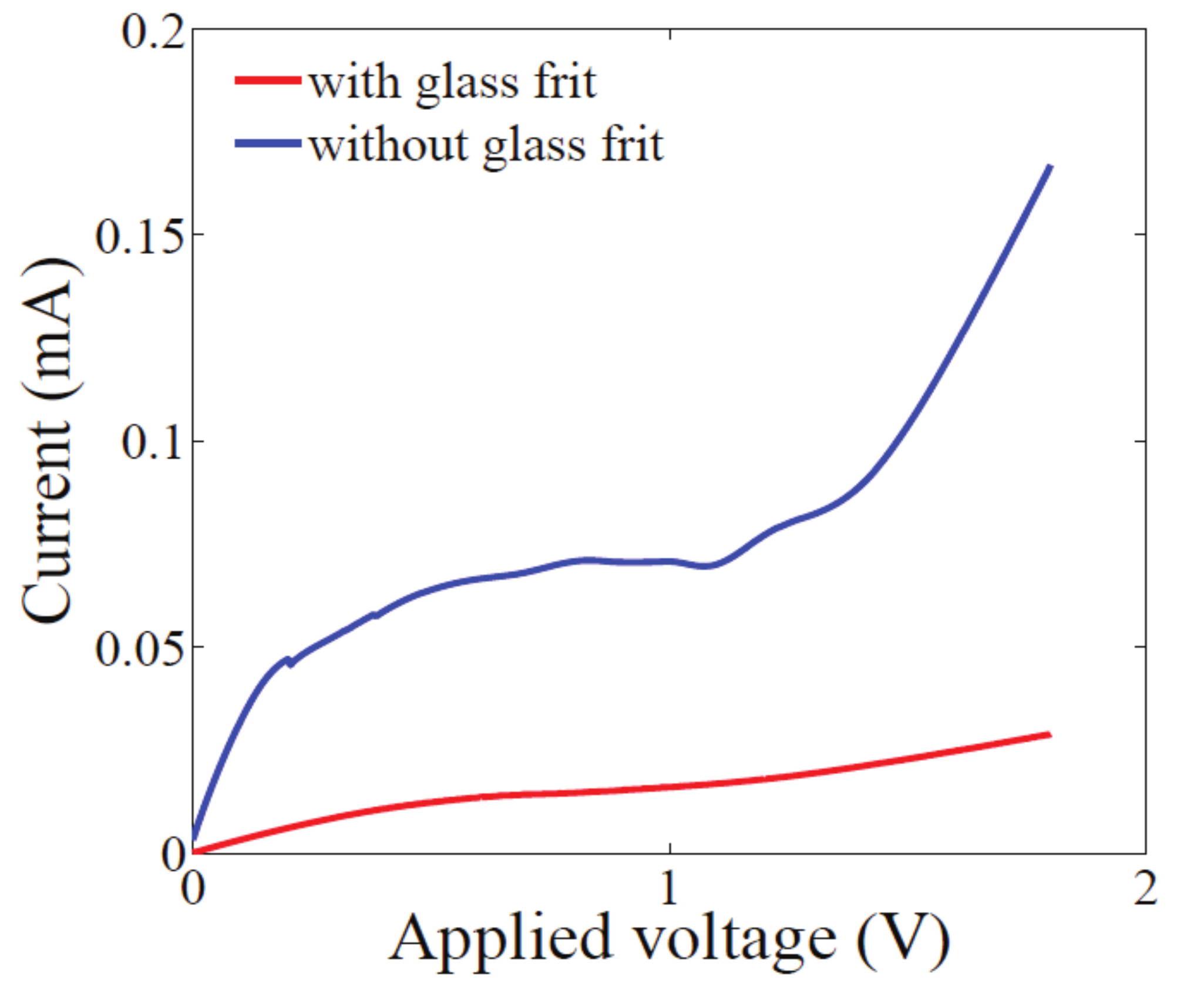}
\caption{Comparison of current-voltage curves with and without a glass frit in a $0.1$ mM CuSO$_4$ solution, where electro-osmotic instability is suppressed in the presence of a glass frit.}
\label{fig:SFnoglassfrit}
\end{figure}

To substantiate our observations, we measured and compared the current-voltage curves for a $0.1$ mM CuSO$_4$ solution with and without a glass frit in the device, as presented in Fig. \ref{fig:SFnoglassfrit}.  The current in the solution without the glass frit is higher than in the device with the glass frit. This is expected since the porous material provides significant resistance to diffusion. We can also use this additional measurement to check the validity of our parameter fitting of the series resistance, $R_s$, in the main text, which is dominated by reservoir resistance, $R_{res}$, at low salt concentration.  In Fig. 5e, we inferred $R_{res}$=5 $k\Omega$/mm for the 0.1mM solution with a 3mm thick reservoir.  Here without the glass frit, the reservoir length (electrode separation) is now 4mm, and the slope in the new figure below gives a similar Ohmic resistance (inverse slope at low voltage) of 3 $k\Omega$/mm.

Without the glass frit, we can also confirm the well-known current-voltage behavior of a membrane in contact with free solution (black curve). The current dramatically increases above a critical voltage of about 1.3 V, well into the limiting current regime (plateau). This feature indicates the occurrence of electro-osmotic instability and is consistent with previous literature results for the very same experimental setup (Nafion with copper sulfate solutions) [1]. With the glass frit in place, this dramatic current increase disappears at higher voltages. Electro-osmotic instability is suppressed in the glass frit, with pore sizes down to sub-micron levels.  This supports our inference that the observed over-limiting current with a glass frit is due to surface conductance and/or electro-osmotic flow effects, along with2 the quantitative comparisons to theoretical predictions shown in the main text.

\subsection*{4. Chronoamperometry measurement during the deionization extraction}
\begin{figure}[h]
\includegraphics[width=0.5\columnwidth,keepaspectratio=true]{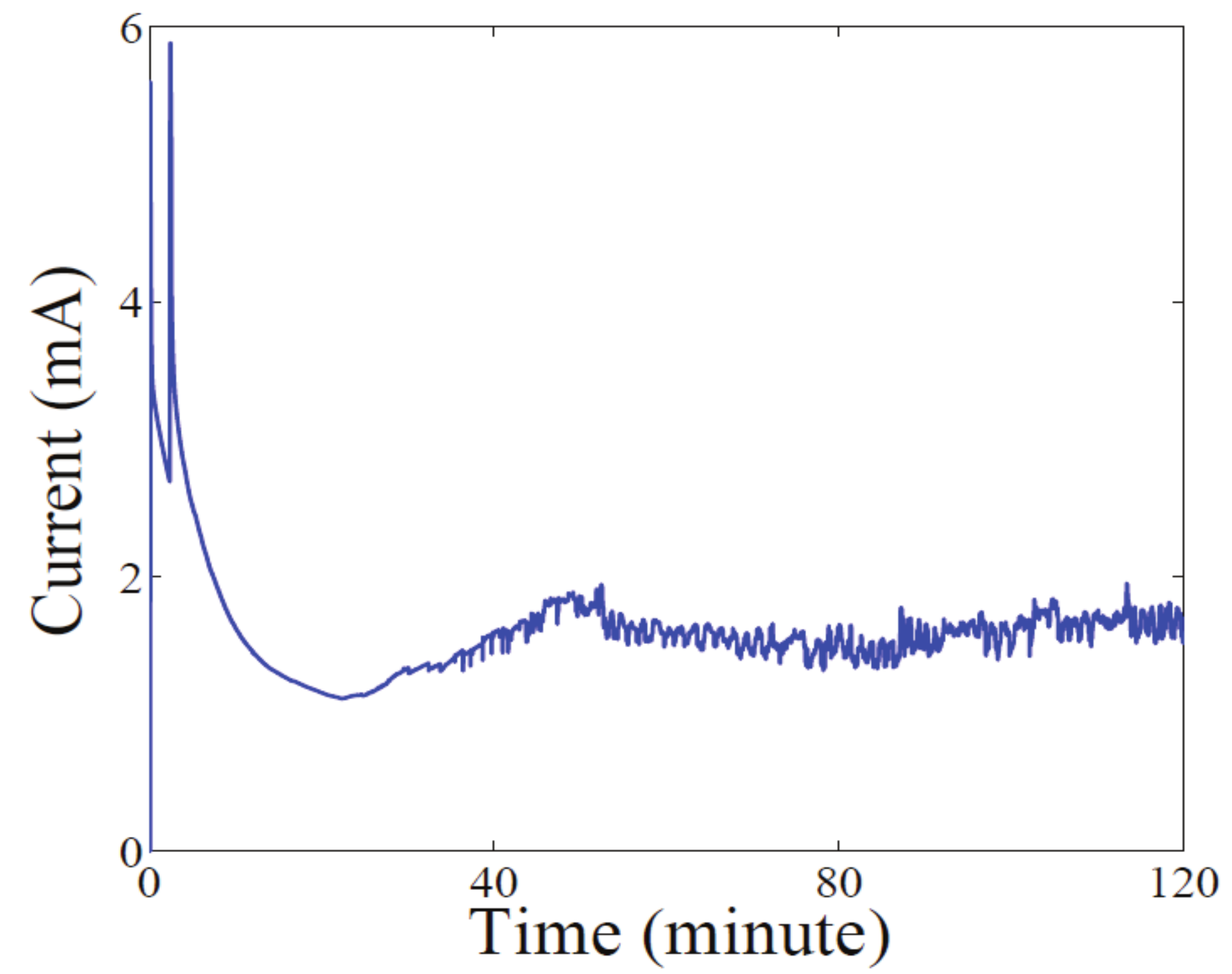}
\caption{Chronoamperometry measurement for a $100$ $mM$ CuSO$_4$ solution at a constant voltage ($1.5$ V) and a slow extraction flow rate of $10^{-2}$ ml/h ( $0.16$ $\mu$l/min). }
\label{fig:SFchoronoamper}
\end{figure}

A constant applied voltage ($1.5$ V) was applied during extraction of the deionization solution.  The chronoamperometry measurement for a $100$  mM CuSO$_4$ solution is presented in Fig. \ref{fig:SFchoronoamper} at a slow extraction flow rate of $10^{-2}$  ml/h ($0.16 \mu$l/min).

\subsection*{5. Impedance measurement for solution conductivity}
\begin{figure}[h]
\includegraphics[width=\columnwidth,keepaspectratio=true]{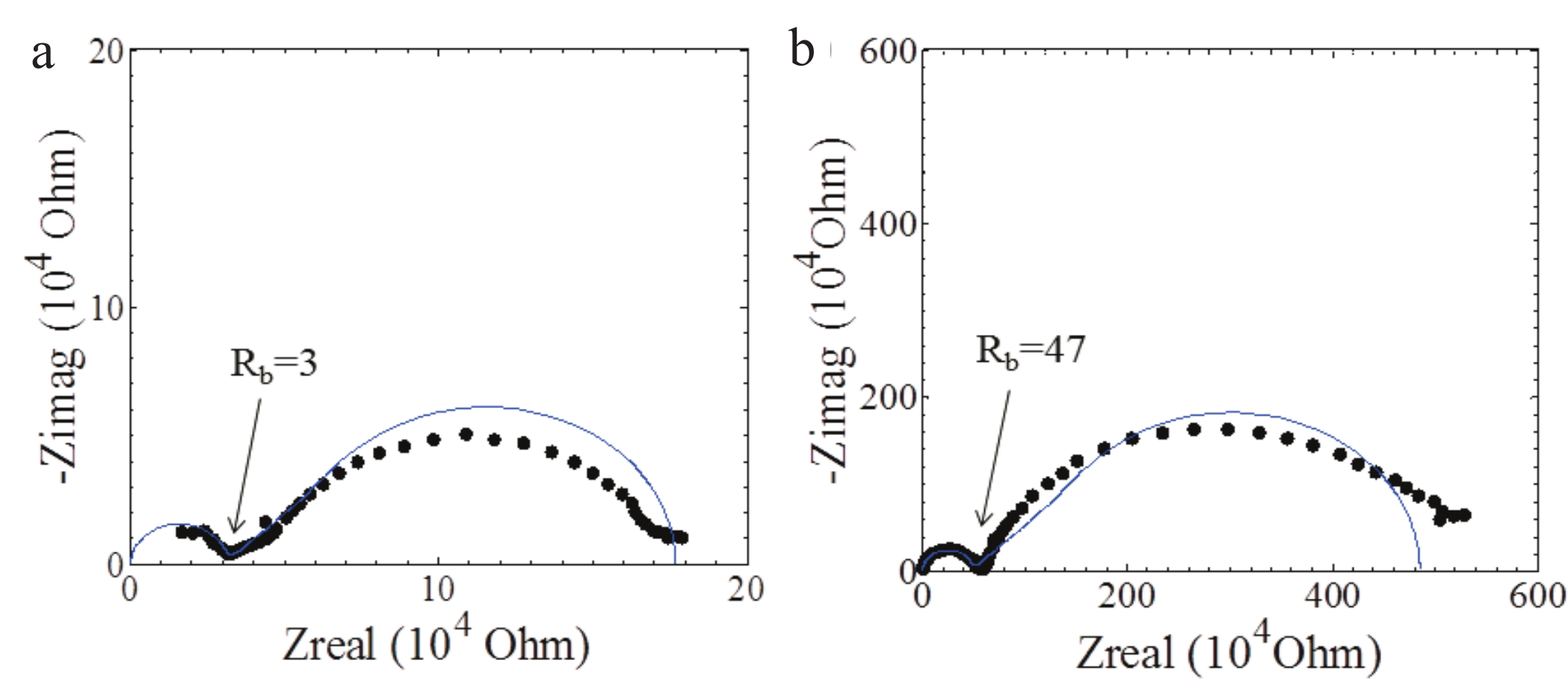}
\caption{(a) Nyquist plot of the impedance spectrum for initial bulk solution of $0.1$ M, and (b) Nyquist plot of the impedance spectrum for the extracted solution at flow rate of $0.15$ $\mu$l/min. Measured data is plotted in points while the line indicates the fitted curve. 
}
\label{fig:supimpedance}
\end{figure}

The copper sulfate solution was infused into a polymer tube with a diameter of $762 \mu$m and a length of $1 cm$. The impedance data, shown in Fig. \ref{fig:supimpedance}, was fit by an equivalent electrical circuit, which is composed of a Warburg element for diffusion in serial with a parallel RC circuit. The solution conductivity was calculated from the fitted bulk resistance.

\subsection*{6. Measurement of pH , calculation of surface charge density, and pK effect}
\begin{figure}[h]
\includegraphics[width=\columnwidth,keepaspectratio=true]{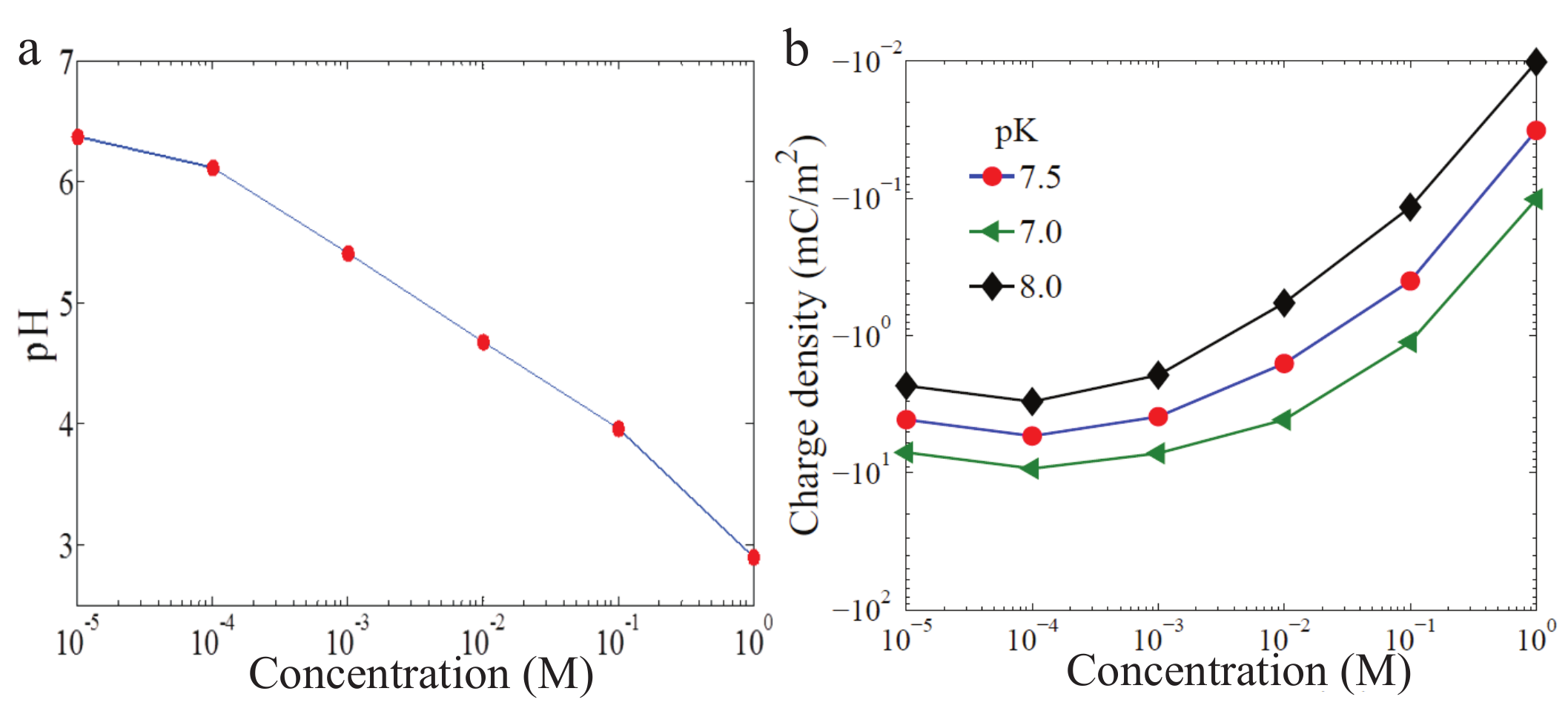}
\caption{(a) The measured pH of copper sulfate solution for various concentrations. (b) The calculated surface charge density dependent on concentration.}
\label{fig:suppHchrge}
\end{figure}
For a binary z:z electrolyte, the distribution is described by the mean-field Poisson-Boltzmann equation, where the effect of overlapping double layers is ignored [2,3]. The surface charge density associated with the diffusion layer potential satisfies the Grahame equation,
\begin{equation*}
\sigma(\zeta)=\frac{2\epsilon\epsilon_0k_BT\kappa}{ze}sinh\left(\frac{ze\zeta}{2k_BT}\right),
\label{eq:surfacechargedensity}
\end{equation*}
Where $e$, $k_B$ and $T$ are the electron charge, the Boltzmann constant, and the temperature, respectively; $\kappa^{-1}$ is the Debye screening length, defined by $\kappa^2=\frac{2(ze)^2n}{\epsilon\epsilon_0k_BT}$, where $\epsilon_0$, $\epsilon$, and $n$ are the permittivity of the vacuum, the dielectric constant of water, and the number of density of ions proportional to concentration, respectively.

The diffusion layer potential ($\zeta$) depends on surface charge density as follows,
\begin{equation*}
\zeta(\sigma)=\frac{k_BT}{e}\left[\text{ln}\left(\frac{-\sigma}{e\Gamma+\sigma}\right)+\text{ln}(10)(pK-pH)\right]-\frac{\sigma}{C},
\label{eq:surfacechargepotential}
\end{equation*}
where $\Gamma=8$ $\text{nm}^{-2}$ is the surface density of chargeable sites, $C=2.9$ $F/\text{m}^2$ is the Stern layer's capacity related to the structure of the silicate-water interface, $pK =7.5$ is the disassociation constant of the silanol group at the silica-water interface ($SiOH\rightleftharpoons SiO^{-}+ H^{+}$), and pH, which is concentration dependent (see Fig. \ref{fig:suppHchrge}a), is measured using a pH meter. The surface charge density was obtained by solving the two coupled equations shown above. 

To further take into account the uncertainty of the pK for silica, we compare the surface charge density for different pK values, as shown in Fig. \ref{fig:suppHchrge}b. Decreasing the pK from from 7.5 to 7.0 increases the surface charge density for all ion concentrations, doubling or tripling in value. Using an enhanced surface charge density does not alter our conclusions and is consistent with the theory that electro-osmotic flow is dominant over surface conduction in our experiments. Furthermore, using an enhanced surface charge density reduces the discrepancy in overlimiting conductance between the simple electro-osmotic flow calculation and the experimental measurements shown in Fig. 3f in the main text.

REFERENCES 

[1]  S. M. Rubinstein,G. Manukyan, A. Staicu, I. Rubinstein,B. Zaltzman, R. G. H. Lammertink, F. Mugele and M. Wessling. Direct observation of a nonequilibrium electro-osmotic instability. \emph{Phys. Rev. Lett.} \textbf{101,} 236101 (2008).

[2] Behrens, S. H. \& Grier, D. G. The charge of glass and silica surfaces. \emph{J. Chem. Phys.} \textbf{115,} 6716--6721 (2001).

[3] van der Heyden, F. H., Stein, D., and Dekker, C. Streaming currents in a single nanofluidic channel. \emph{Phys. Rev. Lett.} \textbf{95,} 116104 (2005).

\end{document}